\documentclass[sn-basic]{sn-jnl}
\jyear{2023}%
\usepackage{amsmath,amsthm,amsfonts}
\usepackage{amssymb, enumitem}
\usepackage{graphicx}
\usepackage{placeins}
\usepackage{epsfig}
\usepackage[caption=false]{subfig}
\usepackage{dblfloatfix}
\usepackage{color}

\theoremstyle{thmstyleone}%

\theoremstyle{thmstyletwo}%
\newtheorem{remark}{Remark}%

\theoremstyle{thmstylethree}%

\begin{document}

\title[Introduction to Process Control]{Teaching control with Basic Maths: \\ \emph{Introduction to Process Control} course as a novel educational approach for undergraduate engineering programs}

\author[1]{\fnm{Julio Elias} \sur{Normey-Rico}}\email{julio.normey@ufsc.br}

\author*[1]{\fnm{Marcelo Menezes} \sur{Morato}}\email{marcelomnzm@gmail.com}
\equalcont{These authors contributed equally to this work.}

\affil*[1]{\orgdiv{Departamento de Automa\c{c}\~ao e Sistemas}, \orgname{Universidade Federal de Santa Catarina}, \city{Florian\'opolis}, \country{Brazil}}


\abstract{
\textcolor{red}{In this article, we discuss a novel education approach to control theory in undergraduate engineering programs. In particular, we elaborate on the inclusion of an introductory course on process control during the first years of the program, to appear right after the students undergo basic calculus and physics courses. Our novel teaching proposal comprises debating the basic elements of control theory without requiring any background on advanced mathematical frameworks from the part of the students. The methodology addresses, conceptually, the majority of the steps required for the analysis and design of simple control systems. Herein, we thoroughly detail this educational guideline, as well as tools that can be used in the classroom. Furthermore, we propose a cheap test-bench kit and an open-source numerical simulator that can be used to carry out experiments during the proposed course. Most importantly, we also assess on how the \emph{Introduction to process control} course has affected the undergraduate program on Control and Automation Engineering at \textit{Universidade Federal de Santa Catarina} (UFSC, Brazil). Specifically, we debate the outcomes of implementing our education approach at UFSC from 2016 to 2023, considering students' rates of success in other control courses and perspectives on how the chair helped them throughout the course of their \textcolor{red}{program}. Based on randomised interviews, we indicate that our educational approach has had good teaching-learning results: students tend to be more motivated for other control-related subjects, while exhibiting higher rates of success.}}




\maketitle

\section{Introduction}
\label{sec1}
\noindent In undergraduate engineering \textcolor{red}{programs} worldwide, the study of process control (i.e. control systems and corresponding theoretical results, generally referred to as \emph{control theory}) is typically divided in two separate parts:
\begin{enumerate}
    \item The first part usually comes with a more tender approach, focusing on basic topics such as the study of (linear, time-invariant) continuous-time systems. In general, \textcolor{red}{frequency-domain mathematical tools are used} (e.g. Laplace and Fourier transforms, root locus design, etc), \textcolor{red}{for both analysis and synthesis}. \textcolor{red}{Typically}, systems are represented by means of (input-output) transfer function models, \textcolor{red}{c.f.} \citep{franklin2013sistemas, garcia2017controle};
    \item In the second part (with courses usually named as \textcolor{red}{``advanced'' or ``modern''} control), the focus tends to be cast \textcolor{red}{towards} discrete-time systems, sometimes modelled by state-space representations, or also using frequency-domain tools ($\mathcal{Z}$ transforms), see e.g. \citep{ogata1998engenharia,Maciejowski}.
\end{enumerate}

This kind of two-step educational guideline is also be seen in the most popular and traditional control theory textbooks, which are likewise structured in the same fashion, c.f. \citep{Dazzo,astrom1986teaching,ogata1998engenharia,albertos2010feedback,franklin2013sistemas}. \textcolor{red}{In some universities, control theory is divided into analysis and synthesis courses - the first part covering performance metrics, modelling aspects and stability, and the later focusing on controller design.}

\textcolor{red}{In both these educational guidelines}, we observe that the initial syllabus of engineering undergraduate \textcolor{red}{programs} (as well as the mentioned textbooks), regarding control systems, is dedicated to the presentation of calculus tools that serve as theoretical support to the subsequent \textcolor{red}{control analysis and design methods}. \textcolor{red}{Only once these mathematical tools are presented, the diverse features and problems of control systems is actually studied}. Furthermore, this kind of progressive programmatic sequence also enables the simplification of tougher mathematical calculations through specific tools (e.g. the analysis of system responses in the frequency domain instead of convolutional analysis). We stress that, often, comprehensive experimental essays are only possible \emph{once the students are already familiarised} with the mathematical tools for control design, e.g. \citep{cruz2012lego,khan2017teaching}. 

In order to illustrate the detailed context, we make reference to the curriculum of control engineering courses in Brazil: Table \ref{CAEnoBrasilTabela} summarises how many control subjects are included in the curriculum and how early they are taken along the program, \textcolor{red}{considering seven distinct control engineering undergraduate programs in Brazil. In particular, we chose\footnote{\textcolor{red}{We emphasise that the university ranking debated in \citep{morandin2020desempenho} is widely popular in Brazil. Anyhow, we indicate that the Brazilian National Exam of Student Performance (\emph{ENADE}) also offers several qualitative discussions on the topic of educational quality, per country region and per program category (public university, private engineering school, federal institutes, and so forth), which are of scholastic interest.}} programs from qualitatively established universities, taking into account the Brazilian university ranking debated in \citep{morandin2020desempenho}.}

\textcolor{red}{With regard to Table \ref{CAEnoBrasilTabela}, we particularly highlight the fact that the} students from the Control and Automation Engineering (CAE) \textcolor{red}{undergraduate program} at \textcolor{red}{\textit{Universidade Federal de Santa Catarina} (UFSC)} have the \emph{earliest} contact to control subject, due to the educational approach that is detailed in this article. The other university courses usually present control systems in the middle\footnote{We note that the control subjects in the fourth semester of the courses at UFRGS and UFRJ cover, in fact, modelling, system and signal topics, and not really control systems itself.} of \textcolor{red}{their programs} (i.e. in the fifth semester). In Appendix \ref{appendiceACursos}, the Reader can find a detailed description of the control engineering curriculum of these seven surveyed universities.

\begin{table}[ht]
    \centering
      \caption{\textcolor{red}{Undergraduate courses of Control engineering in Brazil}: How many control courses? How early? \\(In all cases, the total number of semesters is ten.)}
\label{CAEnoBrasilTabela}
    \small
    \begin{tabular}{|p{0.3in}|p{1.1in}|p{1in}|p{1.3in}|}\hline
        \hline
        $\#$ & University & Number of control systems courses & Semester of the first control-related course \\ \hline
        $1^{\text{rst}}$ & Universidade Estadual de Campinas (UNICAMP) & $7$& $5^{\text{th}}$ semester \\ \hline
    $2^{\text{nd}}$ & Universidade Federal de Minas Gerais (UFMG) & $6$& $5^{\text{th}}$ semester \\ \hline
    $\mathbf{3}^{\text{\textbf{rd}}}$ & \textbf{Universidade Federal de Santa Catarina (UFSC)} & $8$& $\mathbf{3}^{\text{\textbf{rd}}}$ \textbf{semester } \\\hline
         $4^{\text{th}}$ & Universidade Federal do Rio Grande do Sul (UFRGS) & $9$& $4^{\text{th}}$ semester  \\ \hline
          $5^{\text{th}}$ & Universidade Federal do Rio de Janeiro (UFRJ) & $7$ & $4^{\text{th}}$ semester  \\ \hline
           $6^{\text{th}}$ & Universidade Estadual Paulista (UNESP) & $7$ & $5^{\text{th}}$ semester  \\ \hline
            $7^{\text{th}}$ & Universidade Federal de Pernambuco (UFPE) & $5$ & $6^{\text{th}}$ semester  \\  \hline \hline
    \end{tabular}
\end{table}

We highlight that this ``\emph{traditional}'' \textcolor{red}{educational approach to} control theory has a major disadvantage: the \textcolor{red}{focus is} often centered on the mathematical tools themselves and, therefore, student attention to important theoretical concepts \textcolor{red}{fades}. Moreover, the Authors' over thirty years of classroom experience teaching feedback control to undergraduates indicates a significant problem that arises with the mentioned teaching structure: students tend to use advanced maths in a rather agile (and almost automated) way, while neglecting (by not being aware or not mastering) the important \textcolor{red}{theoretical notions} that support the use of the calculus framework.

\textcolor{red}{Let us briefly illustrate this discussion with two typical situations that are recurrently observed in classroom: 
\begin{itemize}
    \item When analysing given transfer function $G(s)$, we students are requested to compute the static gain. A considerable amount of undergraduates tends to assert, without consistent discernment, that $G(0)$ gives the static gain, simply disregarding the analysis of the conditions that are necessary for the static gain to be, in fact, calculated like so. That is, they directly employ the use of related maths (in this case, the \emph{final value theorem}), even if $G(s)$ represents an unstable system. Moreover, when asked to explain the physical meaning of a process' static gain, students often display difficulties to associate the theoretical concept with the mathematical counterpart (the parameter of the process model).  
    \item Another topic for which students also fudge is the analysis of integral action effects in stable closed-loop schemes, considering tracking of step signals (or the rejection of constant load disturbances). The vast majority of students usually assess this topic by means of steady-state analysis of the resulting closed-loop transfer functions, again without evaluating the necessary hypotheses that should be satisfied or pondering over which type of control action is used in order to guarantee offset-free tracking error. It is not evident to students, most of the time, that one does not need to resort to Laplace domain calculus, for instance, in order to demonstrate that the tracking error converges to the origin in steady-state regime.
\end{itemize}}

From classroom practice, we observe that the teaching-learning process of the main concepts of control theory is, in fact, very complex and challenging, given that much of the basic maths that support the control techniques tend not to be well thoroughly assimilated by the students at the moment that control courses are \textcolor{red}{taught}. \textcolor{red}{Accordingly, we indicate two significant} aspects that could be enhanced in the current undergraduate teaching of control systems: 
\begin{enumerate}
    \item The students' tendency to search for answers in mathematical tools related to the topics of study (and, especially, in ``ready-made'' formulas) and not in the concepts themselves;
    \item The difficulty associated with the use of more complex calculus results in partially impairing the assimilation of important control concepts by the undergraduates.
\end{enumerate}

\textcolor{red}{Moreover, considerable levels of student evasion are repeatedly registered in undergraduate control engineering programs in Brazil \citep{da2019educational}. As indicates \citet{meyer2014engineering}, a key factor of dropouts in engineering undergraduate programs relates to poor performances in maths-related courses and the consequent loss of confidence. With regard to this matter, we highlight that the traditional control theory teaching scheme also contributes negatively: students study tougher calculus topics first, subjects with higher rates of failure, without being able to associate these tools with the professional aspect of engineering, thus contributing to non-persisting along the program.}

To the best of our knowledge (and from discussions with colleagues from around the world, \textcolor{red}{and within our scientific societies}), the aforementioned situation is repeatedly observed, \textit{ipsis literis}, in many universities and undergraduate engineering faculties. This issue makes the teaching of control theory arduous, \textcolor{red}{with students exhibiting} persistent study difficulties (mainly in distinguishing between the use of related mathematical tools to the theoretical concepts themselves). In addition, \textcolor{red}{control subjects are usually associated to low student performance, many failures and reduced success rates}.

In this article, we detail a novel educational approach that has been proposed to address and mitigate this problem. In 2016, the curriculum of the undergraduate \textcolor{red}{program} on CAE at UFSC was modified: a new basic control systems course, named \emph{Introduction to process control}, was included in the third semester of the program, to be \textcolor{red}{taught} just after basic calculus (limits, derivatives and integrals) and physics (dynamic equations and laws of motions), alongside intermediate calculus (surface and volume integrals and differential equations). 

The main innovation of this educational approach is that it allows the study of the fundamental concepts of control theory only with basic maths. In such a way, we are able to discuss the leading control techniques applied in industrial contexts: on-off control, proportional control, proportional-integral (PI) control and proportional-integral-derivative (PID) control. The syllabus is integrally centered in time-domain analyses, considering simple first-order models (in both continuous and discrete time). The course also addresses basic notions on control design, with discussions on advantages and limitations of the different strategies. The context of this introductory course is thoroughly detailed in the Authors' textbooks used at in the CAE bachlor degree at UFSC, \citep{normey2021controle,normey2022controle}.

A second important contribution of the proposed educational approach is that, since it is only based on simple maths, the undergraduates are able to use
such well-known tools in order to solve realistic, yet simplified, process control problems. Moreover, these problems can be validated by means of real experiments and essays, which serve to motivate the students to study calculus, once they grasp the importance of mathematics for the solution of real engineering problems. 

\textcolor{red}{In the remainder of this paper, we detail the main ideas of our novel educational approach regarding control systems, considering only basic maths and physics and with focus on the intuitive notions of control. In particular, the following sections are ordered with respect to structure of the \emph{Introduction to process control} syllabus\footnote{\textcolor{red}{Complementary, the full syllabus of the proposed \emph{Introduction to process control} course can be found in Appendix \ref{appendixSyllabus}.}}: In Section \ref{sec2}, we detail how to discuss control theory to uninitiated students, considering the concepts of what is a process, what are actuators and sensors, manipulated and control variables, disturbances, and block diagram representations. In Section \ref{sec3}, we elaborate on the presentation of system models in continuous- and discrete-time, as well as how to infer on system characteristics. In Sections \ref{sec4} and \ref{sec5}, we present the teaching approaches to on-off and proportional control, respectively. Then, in Section \ref{sec6}, we elaborate on how to discuss the idea of using integral action in closed-loop, which leads to the debate on PI and PID control provided in Section \ref{sec7}.} Of significance importance to the educational approach, we present, in Section \ref{sec8}, a cheap experimental test-bench that can be used to validate and demonstrate the aforementioned topics. Moreover, in Section \ref{sec8b}, we present an open-source simulation software, based in \textcolor{red}{Python}, that can be used to validate theoretical concepts.

In Section \ref{sec9}, we debate how the \emph{Introduction to process control} course has positively affected the overall \textcolor{red}{undergraduate} experience and the obtained teaching-learning results from 2016 to 2023. Specifically, we provide discussions in the sense of how the course enabled students to perceive the key control concepts \emph{before} studying the related advanced mathematical tools. We also show how this educational approach has helped students to perform better in other, more advanced, control courses. Concluding remarks and overall perspectives are presented in Section \ref{secconc}.

\textcolor{red}{
\begin{remark}
Along the following Sections, we detail how to teach and elaborate on different topics of control theory to undergraduate students at their very early semesters of the program. With respect to this matter, we emphasise that our discussions are based on the use of (continuous, and discrete, linear and nonlinear) \textbf{first-order} system models. Accordingly, many of the demonstrations and assessments provided in our educational guideline are only valid for these kinds of models (i.e. computation of closed-loop settling time, steady-state error, and so forth). Nevertheless, as argued in the prequel, our focus is in teaching the \textbf{key concepts and notions} of control theory, which will be later exploited with more advanced mathematical tools in consecutive courses.
\end{remark}
}

\section{Discussing process control}
\label{sec2}

\noindent In the following sections, we detail the proposed educational approach to teach control to undergraduate engineering students. Accordingly, we begin by discussing \textcolor{red}{how to} introduce the first notions of control systems.

\textcolor{red}{From a conceptual point of view, the first idea to be discussed how a process as can be understood as a system that transforms (or modifies) some property of a material or element, potentially converting it into a new product. The second key idea is that, when we talk about process \emph{control}, one requires to be able to act upon the process, as well as to observe the property that one wants to modify. Accordingly, the concepts of manipulated variable (or control input) and controlled variable (or output) naturally arise. At the same time, the concept of disturbance, or a variable that affects the process but cannot be manipulated, can also be introduced. For most real-world processes in practice (which can operate coherently without a proper control system), if  disturbances were not present, it would suffice to choose an appropriate value for the control input to keep it operating with a certain desired output behaviour. None of these concepts require any mathematics to be explained, and many day-to-day examples can be used to illustrate these ideas.}

The concepts of open-loop control and closed-loop control are also fundamental and can be explained from the beginning of the course, as well as the concepts of manual and automatic control operations. 

Discussing operational specifications for the process is another concept that can be introduced intuitively: \textcolor{red}{Why and do we want to control a given process? How do we want to control it?} What are our objectives? At this point, it is important to mention aspects related to the control layers typically found in the industry, aiming to present the general scope of the main issues in a process before delving into specific (local) problems.

Finally, another important concept to introduce early on is related to the operational characteristics of the process under study, defining operating ranges for each of the variables involved. It is common, when studying control systems using only local mathematical models, to forget about these practical details and work with incorrect (infeasible) values of the variables, thereby losing the relationship between the model and the real process. Drawing an operating map of the process, considering the control variables, disturbances, and output, is essential for the student to understand the problem they are analysing.

\section{Presenting models and system characteristics}
\label{sec3}

\noindent The concept of a model is a fundamental topic to be introduced in a basic course. The undergraduate should understand that once it is intuitively established that control systems can be employed to achieve a specific operating condition in a process, it becomes necessary to find a systematic procedure for defining these control actions. For this purpose, the need to study the process operation is highlighted, and the simplest way to do so is through a mathematical description of the associated phenomena, using models.

Associating mathematical models with simple everyday problems is an important motivation for students. By using basic mathematical and physical tools from the early stages of engineering undergraduate courses for these studies, it also serves as an incentive to appreciate these disciplines. We emphasize some important aspects to be taken into consideration in our proposed educational approach:
\begin{itemize}
    \item We don't need complex equations to describe and explain the fundamental concepts of dynamic systems; we can use only first-order models and static models;
    \item Such models should not be limited to being linear, as this could give the false impression that we can fully ``\emph{represent the world}'' by means of linear equations. It is important to introduce and explore nonlinear models as well to account for the complexities and nonlinearities present in many real-world systems. This helps students understand the limitations of linear models and appreciate the need for more advanced modelling techniques when dealing with complex systems;
    \item We should use both continuous-time and discrete-time models to demonstrate the generality of the theory and analyse various processes (including stable, unstable, and integrator systems);
    \item Static models are indeed useful for illustrating situations where the dynamics of a system can be considered instantaneous. For example, in many cases, actuators and sensors can be treated as \textcolor{red}{static} systems, as their response times are much faster than those of the processes they are respectively connected to. We stress that by using static models, we can simplify the analysis and focus on the steady-state behaviours, including those of nonlinear plants.
    \item The concept of stability can indeed be introduced intuitively without the need for a more formal and rigorous theoretical presentation:
    \begin{itemize}
\item To illustrate stability intuitively, examples can be used to demonstrate different scenarios. For instance, the teacher can ask the undergraduates to consider a ball placed at the bottom of a bowl: if the bowl represents a stable operation point, the ball will roll back to the bottom whenever it is perturbed or displaced slightly. Through this example, the teacher can emphasise how stability is related to processes whose behaviours tend to return to an equilibrium state;
\item By using relatable examples and emphasising the concept of returning to equilibrium (or a bounded behaviour within a given operational range), students can develop an intuitive understanding of stability without any formal mathematical analysis and the related stability criteria.
\end{itemize}
\end{itemize}

Next, we present two interesting models that can be used in order to illustrate the ideas debated in the prequel: (\textit{i}) a continuous-time system: the speed regulation of a vehicle in motion; and (\textit{ii}) a discrete-time process: the temporal evolution of a debt. Other simple models that can be considered are temperature control of an oven, level control of a tank, and balance tracking in a savings account.

Consider a car in traffic with a velocity $v(t) \in [0, ~v_{\text{max}}]$ ($t \in \mathbb{R}$) on a road with an inclination $\theta(t) \in [-10^o, ~10^o]$. Using Newton's Second Law with respect to the axis parallel to the road, we obtain the following description:
\begin{eqnarray}
m\frac{dv(t)}{dt} &=& K_m u(t) - K_a v^2(t) - mg \sin (\theta(t)) \text{,}
\end{eqnarray}
\noindent where $m$ is the mass of the car, $K_a$ is the air friction constant, and $K_m$ is the motor constant, which relates its force to the accelerator input signal $u(t)$ (considered between $0$ and $100$ \%).

The model that represents the monthly evolution of a debt $D(k)$ in month ($k \in \mathbb{Z}$), with an initial value $D(0)=D_i$, and with a fixed interest rate $j$ and monthly payment $P_m(k)$ equal to a fixed fraction $x$ of the previous month's debt ($j < x < 1$), can be written as:
\begin{eqnarray}
D(k) &=& D(k-1) + j D(k-1) - P_m(k) + A(k) \, \text{,} 
\end{eqnarray}
\noindent with $P_m(k)\,:=\,x D(k-1)$ and $A(k)$ representing an additional credit increase that may or may not be requested in month $k$. The reduced model of this system is:
\begin{eqnarray}
D(k) &=& aD(k-1) + A(k) \, \text{,} 
\end{eqnarray}
\noindent with $a\,:=\,1+j-x$ satisfying $a\,<\,1$. In this system, we can control the value of the debt by modifying the additional credit requests, with $A$ and $D$ limited to the interval $[0, D_{\text{max}}]$.

By using simple models like these, we can introduce the concept of the operating range of variables, study static characteristics, explain the concepts of static gain and time constant, and observe the effect of disturbances on controlled dynamics.

In the case of the car, the static model is clearly nonlinear, which also allows us to explore the concept of linearity (which is already familiar from algebra courses). For example, for a flat road ($\theta(t) = 0$), the static relationship between $u$ and $v$ is given by the nonlinear relationship $v \,=\, \sqrt{\left(\frac{K_m}{K_a}\right)u}$. By considering this function only within the physical range of the variables (accelerator position between $0$ and $100$ \%, and velocity between $0$ and $v_{\text{max}}$, the maximum car velocity), we obtain one of the curves of the car's static characteristic (the others can be calculated in the same way for different incline values). In the case of the debt, the static model leads to $D = \frac{1}{1-a}A$, which is a linear relationship, defining the static characteristic as a straight line passing through the origin. The static characteristic is a fundamental tool for students to understand the operation of processes at different operating points.

The property of stability can be introduced in a simple manner and applied to examples. For this purpose, we consider a given process operating at a certain operating point. \textcolor{red}{We assume} that a variation in the control signal (or disturbance) of finite amplitude is introduced to the system during a finite time interval. After this interval, the signal no longer affects the system. The process variable will change over time, but if it returns to the equilibrium point where it was initially after a finite time, we can say that this process exhibits stable behavior at that operating point\footnote{The experiment must respect the established \textcolor{red}{bounds} for the considered variables.}.

For stable operating points, the concept of static gain arises in a simple and intuitive way, defined as the ratio between a small variation in control (in the case of the vehicle system, the throttle input $u$) and the resulting variation in the output (in this case, the velocity) when starting from a given operating point. It becomes clear, then, that the static gain is the slope of the tangent line to the static curve at the chosen operating point. The same idea can be applied to the relationship between perturbation and process output.

\textcolor{red}{The concept of a system's dynamic response is the next that is discussed, which involves the debate on how a process can be driven from one operating point to another. Accordingly, we benefit from the valuable use of simulation in order to help students to analyse the behaviour of a process model over time - and, thus, verify that it aligns with preconceived expectations (from the model's static characteristics). We emphasise some aspects regarding this topic:
\begin{itemize}
    \item Using numerical simulation for discrete-time systems has an advantage in the sense that it is straightforward to translate difference equations (models) into recursive loops (in code);
    \item For the case of continuous systems, we benefit from the use of Euler's derivative approximation to retrieve an approximated recursive loop for continuous models. That is, we indicate that the student can replace the derivative term in the continuous model by a discrete-time difference given a small discretisation time step $T_c$ and an integer sample indicator $k \in \mathbb{Z}$, i.e. $T_c\frac{dx(t)}{dt} \approx x\left((k+1)T_c\right) -x\left(kT_c\right)$. Accordingly, we analyse how any first-order continuous model in the form $\frac{dx(t)}{dt} \,=\, f(x(t),u(t))$ can be approximately represented by a discrete-time model $x(k+1) \,=\, x(k) + T_cf(x(k),u(k))$, for $t\,=\,kT_c$ and a coherent $T_c$.
    \item Thus, by means of numerical simulation, we can explore the concepts of what is a transient response and how to characterise a system's time constant, by observing that different model parameters yield system responses that take different time periods for transitioning from one operating point to another.
\end{itemize}}

Furthermore, it is essential to introduce, at this stage of the curriculum progression, the concepts of sampling, interpolation, and quantisation. This allows students to analyse \textcolor{red}{analog}-to-digital and digital-to-\textcolor{red}{analog} communications, which are necessary for the use of discrete-time control systems in continuous processes. It is relatively simple and intuitive to present these concepts exclusively in the time domain, illustrating them through simulated and experimental examples, such as the choice of an appropriate sampling period for a given process.

For a general analytical solution, we need to resort to solving differential or difference equations (which, in this approach, are limited to first-order systems) for specific scenarios defined by controls and disturbances. Thus, the concept of linear approximate model emerges as an interesting way to find the solutions to these equations. The use of models that approximate small variations around an operating point can be well justified in practice. By employing simple techniques of \textcolor{red}{approximating a nonlinear function by a first-order Taylor polynomial at the operating point}, we can derive incremental linear models that represent the dynamic behaviour of the variables involved in the process near the chosen operating point.

From these approximate linear models, the analytical solution of the first-order differential equation:
\begin{eqnarray}\label{linsys1}
\tau \frac{dy(t)}{dt} + y(t) &=& K_eu(t) + K_q q(t) \, \text{,}
\end{eqnarray}
or the first-order difference equation:
\begin{eqnarray}
y(k) &=& a y(k-1) + b u(k-1) + c q(k-1) \, \text{,}
\end{eqnarray}
\noindent allows for the analysis of the dynamic behaviour of the considered process\footnote{In these models, $u, \,y$ and $q$ represent the incremental variables, with $q$ being the system disturbance.}. 

In the discrete-time case, the solution can be presented to students in a straightforward manner using basic mathematics, employing concepts of geometric progressions. This can demonstrate, for example, how the parameter $a$ is associated with the process's time constant for the defined region.

For the continuous case, it is not necessary to resort to the formal solution of differential equations. An interesting strategy for discussing this topic is to observe the different behaviours of real process variables and associate them with certain types of functions over time. Subsequently, the validation of these functions as actual solutions for first-order models can be verified.

The case of controlling the level of a tank can be used to discuss this topic. By conducting an experiment in which we observe the variation of the tank level without water inflow and with a fixed valve opening for outflow, we will notice that the level decreases over time and always has a negative derivative. However, we will also observe that the rate of decrease (its derivative) is greater at the beginning of the experiment than towards the end. From this, we infer that the derivative is approximately proportional in magnitude to the level, suggesting the use of a decreasing exponential function of the form $y(t) = e^{-t/\tau}$ as a representation for the analysed phenomenon. Through simple inspection, we can subsequently confirm that this function is indeed a solution for the first-order model of this system, and through some manipulations, we can obtain the complete mathematical solution of the equation in a straightforward manner. At this point in the progression, we can already associate the concept of time constant with the parameter $\tau$ in the found solution.

With the tools and concepts of modelling and process behaviours, we can progress with the study of different control systems, which are presented sequentially, in the following sections of this article.

\section{On-off control}
\label{sec4}

\noindent The first control technique that we believe should be introduced in a basic control course is the \emph{on-off} approach (sometimes referred to as \emph{bang-bang} control). This type of control structure is one of the most widely used in industry and household appliances, and moreover, it is easily understood and implemented. Despite this, this technique is often overlooked in many control courses and even in many textbooks.

An on-off controller is feedback approach that turns the control action on or off depending on a condition observed in the process variable $y$, which should be maintained within a certain range $[y_{\text{inf}}, y_{\text{sup}}]$. 

Indeed, implementing a control logic of this type is simple, and, complementary, it is important to debate with undergraduates on how processes behave under the action of such on-off controllers. For example, it becomes interesting to discuss the behaviour of room temperature being heated or a refrigerator with an on-off control scheme, in order to engage students with the understanding of everyday devices.

Another important concept that can be introduced alongside this simple control technique is the relationship between the sign of the process gain and the controller gain in a feedback system. For a system with a positive static gain (such as a heater), the control should be set to the $u_{\text{On}}$ condition (maximum value) when the output reaches the minimum value of the band, while it should be set to the $u_{\text{Off}}$ condition (minimum value) when the process variable reaches its maximum value. That is:
\begin{eqnarray}
\label{onoffcont}
u(t) &:=& \left\{ \begin{array}{ccc} u_{\text{On}}, & \text{if} & y(t) \, < \, y_{\text{inf}} \, \text{,} \\ 
u_{\text{Off}}, & \text{if} & y(t) \, > \, y_{\text{sup}} \, \text{,} \\
\text{Maintained}, & \text{otherwise.} &  \, \end{array}\right.
\end{eqnarray}

In the case of systems with negative static gain (such as a refrigerator), we act in the opposite way, using $u_{\text{Off}}$ when the output reaches the minimum value of the range and $u_{\text{On}}$ in the other condition\footnote{The implementation for discrete systems is the same, replacing the continuous real variable $t$ with a discrete integer variable $k$.}:
\begin{eqnarray}
u(t) &:=& \left\{ \begin{array}{ccc} u_{\text{Off}}, & \text{if} & y(t) \, < \, y_{\text{inf}} \, \text{,} \\ 
u_{\text{On}}, & \text{if} & y(t) \, > \, y_{\text{sup}} \, \text{,} \\
\text{Maintained}, & \text{otherwise.}&  \, \end{array}\right. \, \text{.}
\end{eqnarray}

The behaviour of systems under the effect of on-off feedback controllers is simple to describe analytically, considering the simple process models presented earlier. Thus, the characteristics of the final response and switching times can be calculated directly. Regarding this point, it is important to emphasise to students that there is a trade-off between the simplicity of the controller and the performance obtained by this control. 

\textcolor{red}{Regarding these on-off control schemes, it should be emphasised that a significant advantage is that only a relay-type actuator is required for an adequate implementation. Nevertheless, it is key to remark that the process output is always kept fluctuating within the defined operational range, and the resulting transient responses are determined by the open-loop process dynamics.}

The undergraduate student should be aware that, for many processes, this type of control approach can be suitable. On the other hand, the disadvantages of these controllers serve as a starting point for motivation for the use of proportional controllers.

\section{Proportional control}
\label{sec5}

\noindent Indeed, proportional (P) control naturally emerges as a simple alternative to bang-bang scheme, being able to maintain the process output (and, consequently, the control variable) at a specific fixed operating point.

In the proportional control strategy, the control signal is automatically adjusted based on the difference between the desired set-point and the process output. In this paradigm, a reference signal $r(t)$ is needed to indicate the desired operating point for the output, along with a tracking error signal $e(t) = r(t) - y(t)$. Based on the calculation of this error signal, the proportional control action is defined as follows\footnote{In the discrete-time case, $u(k) = K_p e(k)$.}:
\begin{eqnarray}\label{Pactionlaw}
u(t) &=& K_p e(t) \, \text{,}
\end{eqnarray}

Following the same logic as the on-off control strategy, it is discussed with the students that a positive proportional gain $K_p\,>\,0$ should be used for processes with positive gain, and vice versa\footnote{This condition is valid for all types of feedback controllers.}. \textcolor{red}{Furthermore, it should be emphasised that the P control is only able to act \emph{correctly} if the generated control signal is \emph{admissible}, i.e. it respects the systems' saturation constraints $u(k) \, \in \, [u_{\text{min}}, u_{\text{max}}], \, \forall k \, \geq \, 0$. Accordingly, if the error signal is limited within a range called the \emph{proportional band} (PB), it is implied that the corresponding control action is admissible. In particular, the PB defines the range of proportional action for this control strategy, as if the error exceeds this range, the P control behaves like an on-off controller (due to the saturation effects). We have\footnote{Here, we use $u_{\text{max}}$ and $u_{\text{min}}$ as the maximum and minimum allowable values for the manipulated variable, respectively.}:
\begin{eqnarray}
\text{PB} &:=& \frac{\left(u_{\text{max}}-u_{\text{min}}\right)}{K_p} \, \text{.}
\end{eqnarray}}

\textcolor{red}{
\begin{remark}
    Other practical aspects on the effects of saturation are debated, within the proposed introductory control course, \emph{after} PI controllers are explained. Accordingly, refer to Section \ref{sec7}. 
\end{remark}
}

The first advantage of P control over on-off control is that for a constant reference signal $r(t) = r_0$, if the closed-loop system is stable and operates within the operating range of $u$, the process variable will reach a fixed operating point given by $P_0 = (u_0, y_0)$. It is straightforward to calculate that since $y_0 = K_eu_0$, and $u_0 = K_p(r_0-y_0)$, we have $y_0 = \frac{K_eK_p}{1+K_eK_p}r_0$, and also $e_0 = \frac{1}{1+K_eK_p}r_0$.

Therefore, it can be emphasised to the students that we will always have a non-zero error with simple proportional controllers. This fact is intuitive because if the error were zero, we would not have any control action to maintain the control signal at $u_0$.


We emphasise that all this static analysis of the system's behaviour does not require complex calculations and is valid for both continuous and discrete systems.

In terms of dynamics, it is relatively simple to show students that under the influence of P controllers, the model of the closed-loop system is also a first-order system. \textcolor{red}{Thus, in addition to the previously calculated static relationships, we can discuss how the time constant of the closed-loop system, given by $\tau_{\text{CL}} \, = \,  = \frac{\tau}{1+K_eK_p}$, directly depends on the choice of the proportional gain. Accordingly,} for larger values of $K_p$ (in magnitude), we observe faster responses in the closed-loop system. \textcolor{red}{The same argument holds for discrete systems, where we obtain a closed-loop model in the form of $y(k)\,=\,a_{\text{CL}}y(k-1) +bK_pr(k-1)$, with $a_{\text{CL}} = a - bK_p$. Accordingly, larger values of $K_p$ (with the condition $0 \,< a-bK_p \,< 1$) lead to smaller values of $a_{\text{CL}}$ and, therefore, faster responses.}

In the context of an introductory course, it is important at this point to highlight a concept that often goes unnoticed: P control does not alter the dynamics of the system, which will continue to respond with a time constant $\tau$. \textcolor{red}{Instead, it alters the dynamics of the closed-loop system.} What happens in the transient regime, at the initial instants, is that the controller acts on the process with signals of much higher amplitude than those that lead to the desired equilibrium point. This causes the output to approach the reference signal more rapidly, and then the control action is reduced to prevent the output from exceeding $r$ and reaching equilibrium.

\begin{figure}[htb]
    \centering
    \includegraphics[width=0.5\linewidth]{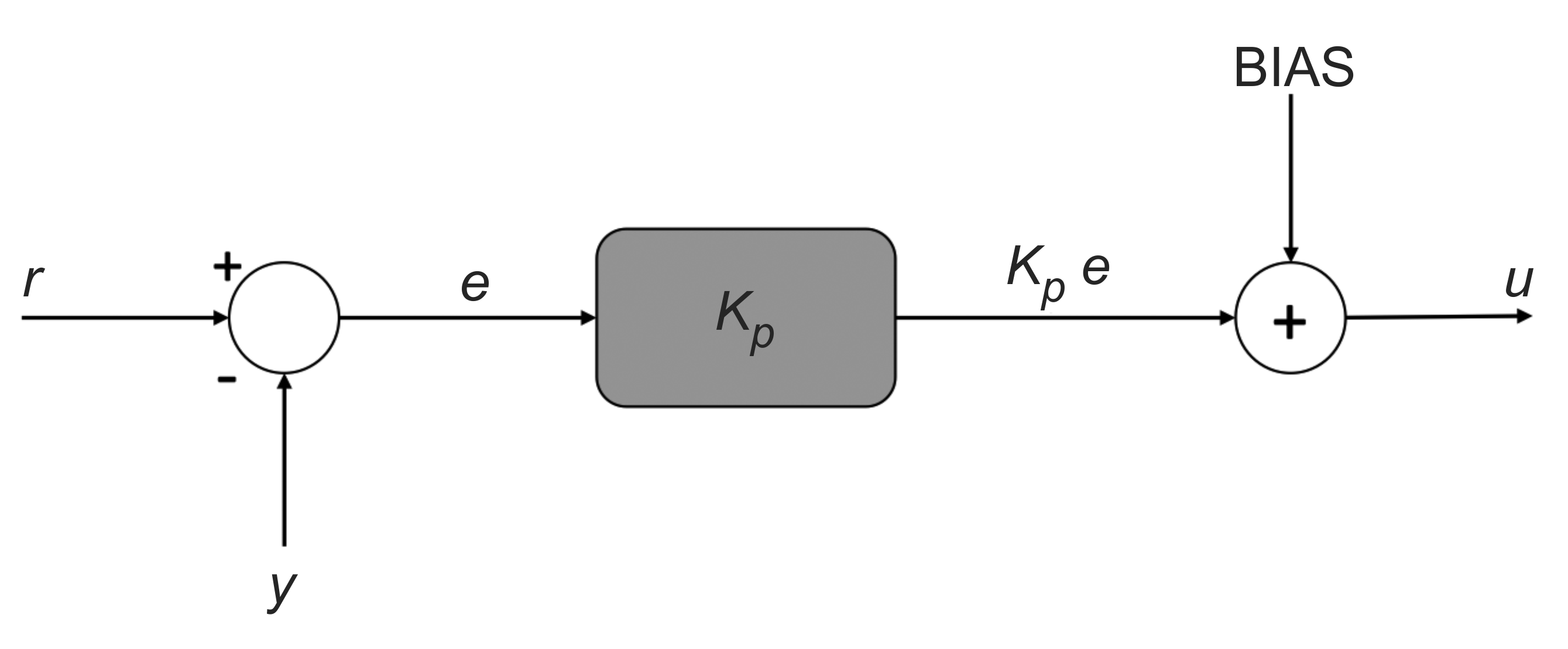}
    \caption{Proportional control implementation with a bias signal.}
    \label{controlePesquema}
\end{figure}

P controllers are very useful for regulating simple processes, and the drawback of not being able to guarantee null error in steady-state can be addressed by adding a supplementary control action, a bias signal (denoted BIAS). Accordingly, the supplemented P control action is given by: $u(t) = K_p e(t) + \text{BIAS}$, as illustrated in Figure \ref{controlePesquema}.

\textcolor{red}{The bias signal is a \textbf{manual} adjustment that aims to impose the desired equilibrium in steady-state. If it is correctly tuned, the process output is steered to the reference signal (i.e., $\lim_{t\to+\infty}y(t)\,=\, r_0$).}

\textcolor{red}{As an example, consider a linear process in the form of Eq. \eqref{linsys1}, subject to a P control action in the form of Eq. \eqref{Pactionlaw} with an additive bias term given by BIAS $=\frac{(y_0-K_qq_0)}{K_e}\,=\,\frac{(r_0-K_qq_0)}{K_e}$. In closed-loop operation, the resulting error signal $e(t)\,=\,r(t)-y(t)$, given $r(t)$ as a step signal with amplitude $r_0$, will converge to the origin and the corresponding P control signal will be given by the bias term when the system reaches the operating point, that is, $\lim_{t\to\infty}u(t)=$BIAS.}

In the discussion of this topic, it should be emphasised to the students that the correct tuning of the control scheme depends on the chosen operating point, and if it changes due to disturbances, the compensation signal (bias term) will also need to be readjusted manually. Furthermore, this approach can also be directly applied to systems described by nonlinear models, for which the tuning of the bias signal is determined based on the static description of the process.

\section{The idea of an integral action}
\label{sec6}

\noindent \textcolor{red}{The main conclusions that can be sustained, considering the use of P controllers, is that offset-free set-point tracking and null-error disturbance rejection cannot be thoroughly enforced. Anyhow, it should be emphasised that the use of an additive bias term can resolve these issues, given that the disturbance and set-point are known.} 

\textcolor{red}{Accordingly, the next (open) question to be debated with students is as follows: can the inclusion of the bias term be \emph{automatic}, without requiring any kind of manual adjustment? From a corresponding debate, the idea of an integral action arises as a simple way to ensure that a closed-loop control system achieves zero steady-state error in tracking reference signals and constant disturbances, provided that the closed-loop dynamics are stable.}

A straightforward way to introduce the concept of integration, without using advanced mathematical tools, is to describe the following integration law:
\begin{eqnarray}
u(t) &=& \int_{0}^{t} e(\mu) d\mu \, \text{.}
\end{eqnarray}

\begin{figure}[htb]
    \centering
    \includegraphics[width=0.9\linewidth]{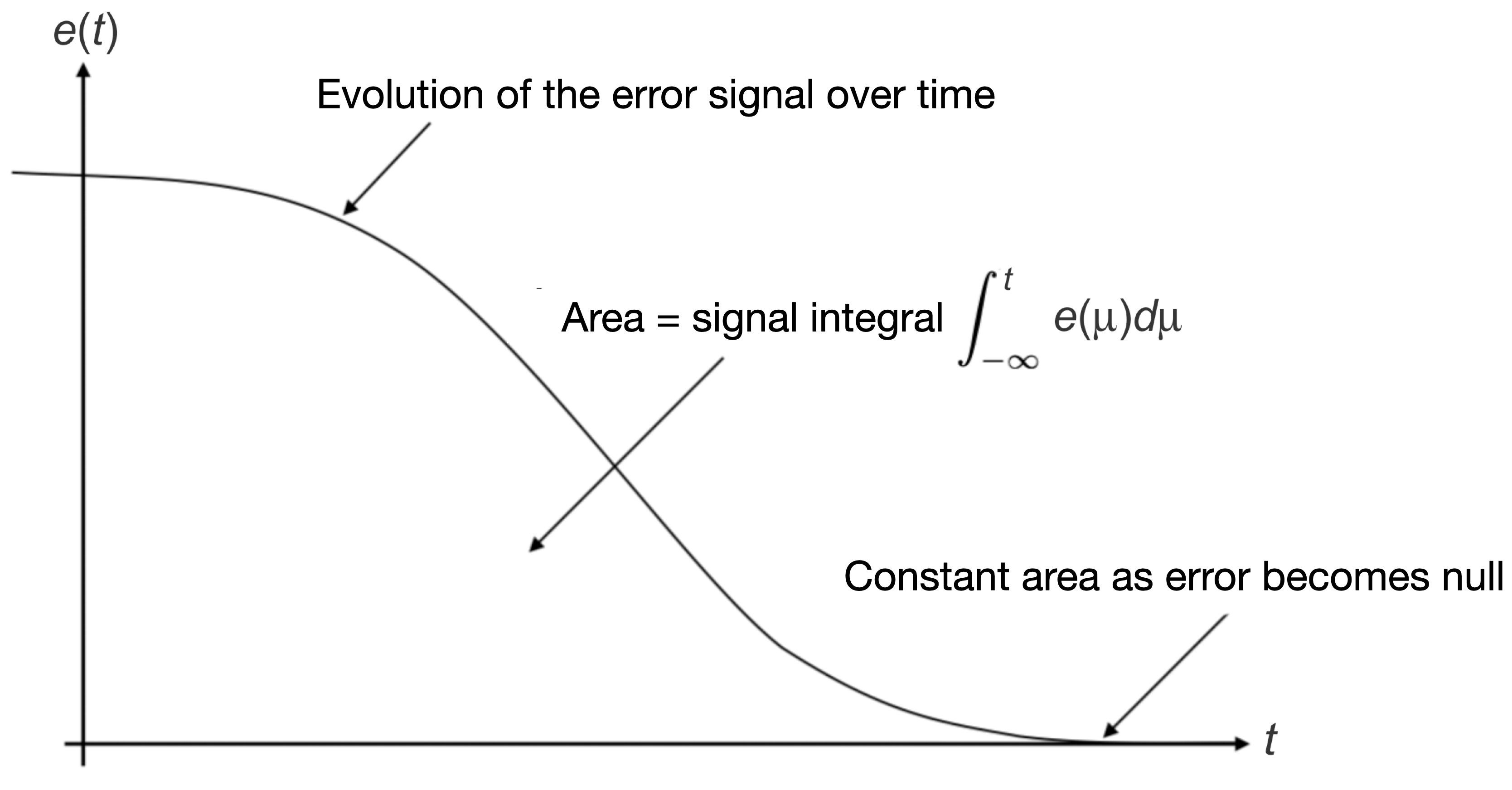}
    \caption{Integral of the error signal.}
    \label{3erroaolongodotempoPI}
\end{figure}

Thus, it is enough to emphasise that the signal $u(t)$ is equivalent to the area under the function $e(\mu)$ for the interval $\mu \in [0, ~t]$. The value of this area, starting from a certain $t'$, will be constant as long as, from that moment on, all values of $e$ are zero, that is, $e(t)=0, \forall t \geq t'$. Therefore, when we introduce an integrator control in a closed-loop system and assume that the system reaches an operating point with constant output $y_0$ and control $u_0$, the tracking error $e_0=r_0-y_0$ will necessarily be zero in steady-state, because otherwise the control signal would not be constant and the system would not be in equilibrium (which represents an evident contradiction). This debate is illustrated in Figure \ref{3erroaolongodotempoPI}, which shows the evolution of an error signal converging to the origin due to the integral effect of the control action.

\textcolor{red}{Using a discrete integral action, give by $u(k) \,=\, T_s \sum_{i=0}^{k} e(i)$, being $T_s$ the sampling time (i.e. the rectangular approximation of the area below the curve, as shows Fig. \ref{3erroaolongodotempoPI}), we can express the action recursively, as follows:}
\begin{eqnarray}
u(k) &=& u(k-1) + T_se(k) \, \text{.}
\end{eqnarray}

Therefore, the condition for the control signal $u(k)$ to remain constant is given by $u(k)=u(k-1)=u_0$, which implies that $e(k)=0$. Once again, with a few mathematical steps, we can show how the integral action ensures zero tracking error.

Thus, a stable closed-loop system under the integral action of a controller has the property that the control signal converges to a value of $u$ such that \textcolor{red}{$\lim_{t\to+\infty}y(t) \,=\,y_0\,=\,r_0$}, for any constant reference signal $r(t) =r_0$ and disturbance signal $q(t)=q_0$, within the operating range of the process.

It should be noted that in practice, the integral control is implemented with an adjustable gain, in order to weigh the integrating action, i.e., $u(t) = K_i \int_{0}^{t} e(\mu) d\mu$ or $u(k) = u(k-1) + K_i T_s e(k)$.

Although the integral control (I) can improve the steady-state performance for constant input signals, the same cannot be said for the transient response. By using past values in the control law, the integral control becomes slower than the proportional control (P), as the relative changes in the control action are small at each sampling period.

Let us analyse, for example, when a process is initialised with $u(0)\,=\,0$ and $y(0)\,=\,0$, with a reference signal $r(k)\,=\,r_0 \,\neq \,y(0)$. Initially, an integral controller will compute the following law: $u(1)\,=\, u(0) + K_iT_s(r_0-0) \,=\, K_iT_sr_0$. Since $T_s$ represents the sampling period, which is usually small, the resulting integral action will have a low amplitude, and therefore, we will observe little variation in the output. This issue can be compensated by using a very large value of $K_i$, but if we increase the control intensity significantly in the initial stages (with values of $u(k)\,>\,u_0$), we will need negative error values to subsequently decrease the value of $u(k)$, leading to oscillatory responses.

It should be noted that, on the other hand, a proportional control uses $u(1)\,=\,K_pr_0$, allowing for a faster action in the initial stages. In this case, there is no problem in using values of $u(k)\,>\,u_0$, as the proportional control \emph{has no memory} and can instantly reduce its intensity as the error changes.

\section{PI and PID control}
\label{sec7}

\noindent Therefore, the programmatic presentation continues with the introduction of PI control. This can be done subtly, taking advantage of the benefits of the integral action in the steady-state and the proportional action in the transient response.
A PI control is a control that performs a weighted sum of the proportional (P) and integral (I) actions. In other words, for an error signal given by $e(t) \,=\, r(t)-y(t)$, we have:
\begin{eqnarray}
 u(t) &=& K_p e(t) + K_i \int_{0}^{t} e(\mu) d\mu \, \text{.}
\end{eqnarray}

In the discrete-time case, we have $e(k) = r(k)-y(k)$ and, thus:
\begin{eqnarray}
 u_P(k) &=& K_p e(k) \, \text{,} \\
 u_I(k) &=& u_I(k-1) + T_sK_ie(k) \, \text{,} \\
 u(k) &=& u_P(k) + u_I(k) \, 
 \end{eqnarray}
\noindent which can be equivalently expressed using:
$$u(k) \,=\, u(k-1) + K_1 e(k) + K_2 e(k-1) \, \text{,}$$ 
with $K_1\,=\,K_p+T_sK_i$ and $K_2\,=\,-K_p$.

When considering the characteristics of P controllers with bias signals, we can observe that the PI control law automatically implements a bias action. Accordingly, PI schemes embed a parallel component to the P action that slowly adjusts to bring the process variable to equilibrium with $e=0$ (and therefore, null P action).

The analytical tuning of PI control for first-order systems, as presented, requires the use of differential and difference equations of second order. In this teaching proposal, we have chosen not to use this approach and instead provide intuitive ideas about the control actions, showing how higher gains result in faster responses but may also introduce oscillations in the closed-loop response, as explained for the I action.

On the practical side, through experiments and simulations, tools can be provided to students on how to adjust the P and I gains. Furthermore, it is worth noting that the literature presents various tabulated methods for tuning these parameters, such as the method by Prof. Skogestad \citep{skogestad2003simple}, which are widely used in the industrial field. Some of these tuning methods are based on empirical studies, such as the well-known Ziegler-Nichols tuning rules \citep{ziegler1942optimum}. Therefore, presenting some of these methods can be done with relative ease, focusing on the following aspects:
\begin{itemize}
\item Which PI configuration is used for each tuning methodology;
\item What are the necessary specifications for each tuning method;
\item What are the control objectives in closed-loop control for each tuning methodology.
\end{itemize}

With this information at hand, the undergraduate student can choose the type of tuning to implement based on the available process model and the alignment of their objectives with the chosen tuning rule.

Indeed, other practical aspects of great importance can be analysed and studied without the need for advanced mathematics, such as the weighting of the reference signal within the proportional action, the use of anti-windup techniques when the control is operating under saturation, and the use of filters on measured signals when they are heavily affected by noise. These topics provide valuable insights into real-world control system implementation and can be explored through practical examples, case studies, and simulations, allowing students to gain a deeper understanding of the challenges and techniques involved in practical control applications.

\subsection{Set-point weighting}

As discussed earlier, in industrial practice, control systems generally aim to reject disturbances as their primary objective. When adjusting a PI controller to achieve a fast response to a disturbance, we must consider that if the reference signal remains unchanged, the control action variation will be generated by the output signal variation ($\Delta y(t)$), and the error variation will be given by $\Delta e(t) = -\Delta y(t)$. Typically, $\Delta y(t)$ will not exhibit an instantaneous variation (as this variation depends on the process dynamics), and therefore, the values of $K_p$ and $K_i$ should be sufficiently large to generate the necessary control signals for fast disturbance rejection.

However, when maintaining the original tuning for $K_p$ and $K_i$, a rapid change in the reference signal (such as a step change) will result in a very high control signal, which can cause peaks in the control action and consequently undesired responses in the system output during the transient regime.

Choosing an intermediate tuning for the PI gains $K_p$ and $K_i$ could be an alternative, seeking an adequate trade-off between rejection speed and peak response to the set-point variations. However, a more elegant (and straightforward) solution is to modify the implementation of the PI controller so that the proportional action, which is responsible for the fast transient response, fully acts on disturbance rejection and only partially on reference changes. This objective can be achieved with the following control law:
\begin{eqnarray}
 u(t) = K_p(\alpha r(t)-y(t)) + K_i \int_{0}^{t} e(\mu) d\mu \, \text{.}
\end{eqnarray}

Note that by using $\alpha < 1$, the proportional action has a lower gain on $r$ compared to $y$. This allows for a higher proportional action when only $y$ varies (i.e., when disturbances occur). This strategy is simple, easy to implement, and significantly improves the performance of the PI control.

\subsection{Anti-windup action}

In practice, control systems in industrial applications typically have a control range limited by the interval $[u_{\text{min}}, u_{\text{max}}]$. If, at a given instant, the computed control action exceeds this range, the actual control applied to the process will be different from the calculated value. This saturation effect has two straightforward consequences that can be explained without advanced mathematics. Firstly, in steady-state, saturation may prevent the system from reaching a desired reference value or rejecting a disturbance if the required control action falls outside the operating range. This is why determining the operating ranges of variables is crucial before designing the control system. Secondly, in transient response, if the control action is limited by saturation, the process variable will evolve more slowly than expected without saturation. These issues affect all types of control schemes (P, I, PI, and PIDs).

However, when integral action is present in the control loop, saturation causes another effect in the transient response known as \emph{windup} (or \emph{integral windup}).

This phenomenon can be explained using a simple example of an integral control law applied to a process with a positive static gain. Consider a system starting from \textcolor{red}{$y(0)\,=\,0$} with a reference signal $r(t) = r_0$ within the acceptable operating range of the process. \textcolor{red}{In steady-state, the system requires a control action $u_0$ for an output $y_0\,=\,r_0$ (assuming that $u_0$ is also within the operating range).} At the initial instant of the transient response, the error is $e(0)\,=\,r_0$. If, at that moment, the control action saturates (\textcolor{red}{for simplicity, we assume that it saturates at the maximum value of the operating range}), we know that the applied control will be $u_{\text{max}}$, while the computed control $u_c\,=\,u(k-1)+K_iT_se(k)$ will exceed $u_{\text{max}}$. Since the applied control is lower than the calculated value, the variation in the output will be smaller than expected, and at the next sampling instant, we will still have a positive error. Then, the control action will be recalculated with an incorrect update of $u(k-1)$ using $u_c$. Since $u_c \,>\, u_{\text{max}}$ and the error is positive, the new calculated control value will be higher than the previous one. However, due to saturation, \textcolor{red}{the controller will continue to apply $u_{\text{max}}$ to the process. This situation effect will remain for several sampling instants, causing the integral action to accumulate. When the output finally reaches the set-point value defined by the reference signal, a change in error sign occurs and, consequently, the integral action starts to decrease.} However, the accumulated value will be much higher than $u_{\text{max}}$, and in order for the control signal to \textcolor{red}{converge to} $u_0$ (the desired value in steady-state), several samples with negative error values are needed to reduce the integral term. As a result, the process response will exhibit a peak, which can be quite high depending on the control and process tuning.

\begin{figure}[htbp]
    \centering
    \includegraphics[width=\linewidth]{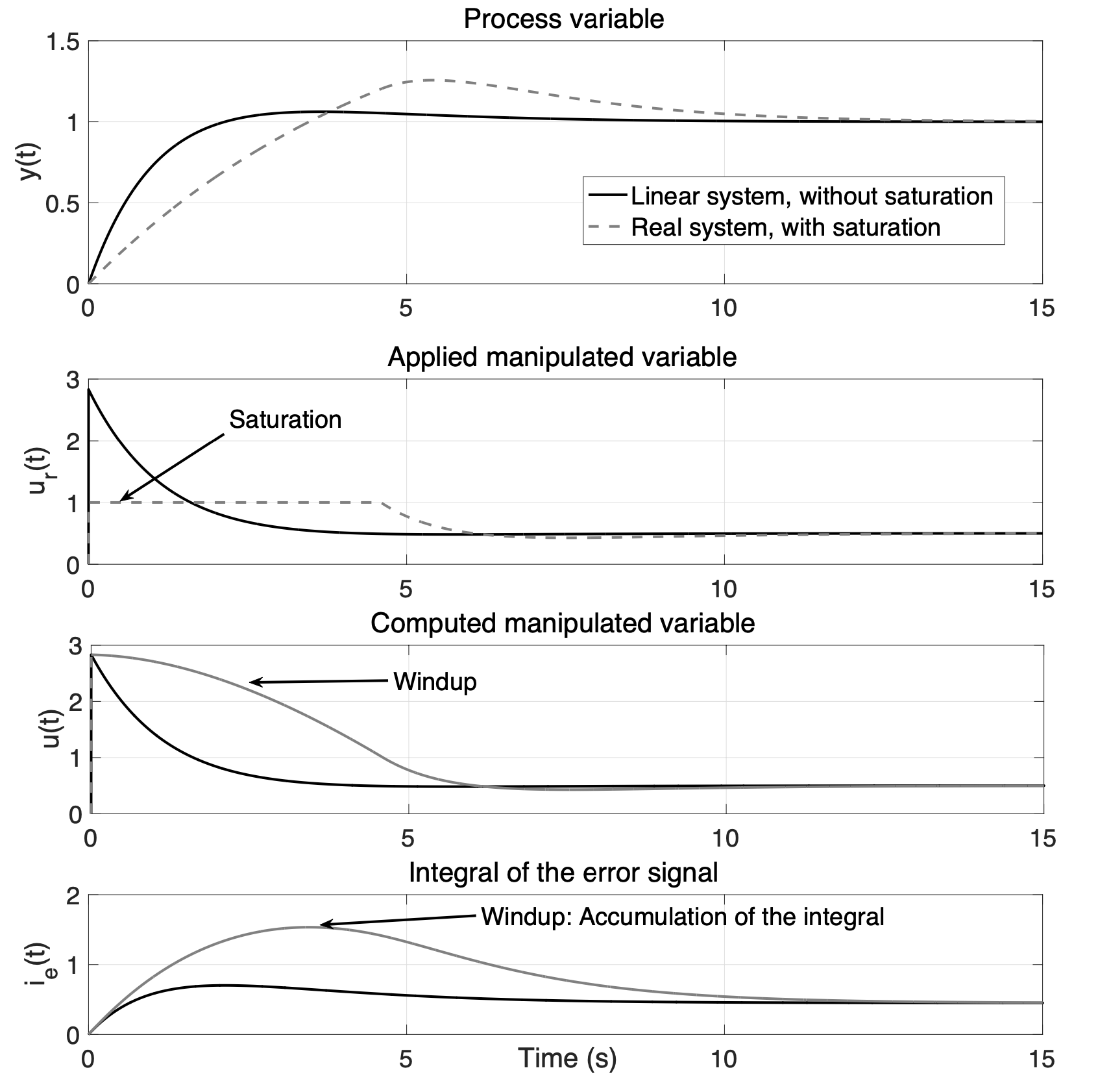}
    \caption{Closed-loop (linear system and PI) responses  without saturation (full lines) and with (dotted lines).}
    \label{saturaI1}
\end{figure}

\textcolor{red}{In order to debate the concept of windup and saturation effects with students, one can consider the example of a simple linear system, with positive gain, regulated by a simple PI controller. The system responses can be analysed without saturation and when saturation occurs, as shows Figure \ref{saturaI1}. Accordingly, by analysing the behaviour of following signals (process output, real applied control action and computed one, before saturation, and the integral of the error), the windup effect can be graphically understood:
\begin{itemize}
    \item The first topic to be highlighted in classroom is that the control range is reduced when saturation happens and, accordingly, we observe \emph{slower} closed-loop responses.
    \item Anyhow, the steady-state control value is not altered, given that, after the transient response, in both cases (with and without saturation), the error integral value will be the same. This is an expected property, given that the steady-state definition $\left(u(+\infty),y(+\infty)\right)$ will depend only on the chosen set-point value, given that it lies within the admissible range.
    \item Analysing the transient response, one should remark to students how the error integral positive part of signal $i_e(t)$ is larger when saturation is implied. Accordingly, its negative part should also be greater (in magnitude), when saturation is implied. As of this, the saturated system response exhibits a larger peak, taking more time to reach the steady-stead - which characterises the windup effect.
\end{itemize}}

In order to avoid this saturation-induced problem, the integral action should be correctly updated so that the control actually applied to the process is used in the recursive calculation. A simple code for this implementation is $u(k)\,=\,u(k-1)+K_iT_se(k)$, with the following update: $u(k) \,=\, u_{\text{max}}$ if $u(k) > u_{\text{max}}$ and $u(k) = u_{\text{min}}$ if $u(k) < u_{\text{min}}$, and updating $u(k)\,=\,u(k-1)$ before the next discrete-time sample. \textcolor{red}{By doing so, during the saturation period, the control signal remains at the limit of the allowed range, allowing the system to rapidly exit the saturation regime when the error signal changes its sign.} This idea can be easily extended to PI control by modifying the equation for the calculation of the control signal $u(k)$.

\textcolor{red}{After analysing how set-point weighting and anti-windup strategies can be used, in the context of PI controllers, the next topics that should be debated with students are derivative control and noise filtering.}

\subsection{Motivating derivative control}

\textcolor{red}{When we apply a step change in the set-point signal for this process, we observe how the controlled variable is steered towards the new set-point value with a certain \emph{speed}.} The \emph{ideal} scenario would be for this process variable variation to be instantaneous, such that the process variable comes very close to the reference and quickly \textcolor{red}{``\emph{slows down}'' in order to reach the desired value without overshoot.}

\textcolor{red}{In order to achieve a response that ``slows down at the right moment''}, it would be beneficial for the controller to know, at each instant, the future error values in order to decide when to start decelerating (i.e. \emph{braking}) the control action. In other words, knowledge of the system's future behavior could assist in constructing an ideal control action, ensuring that the system response exhibits the desired performance.

For this purpose, the instructor can use a simple discrete process model of the form $y(k) = a y(k-1) + b u(k-1)$. Note that for this process, if we desire a certain future behaviour $y(k+1)$ for the output, we can define in advance the desired values of $y(k)$ and calculate the resulting control signal based on the sequence of values for $y$. This could be achieved using the following control law:
\begin{eqnarray}
u(k) &=& (1/b) y(k+1) - (a/b) y(k) \, \text{.}
\end{eqnarray}

Thus, if we knew the future values of the process output (in this example, since the system is first-order, knowing the output value one step ahead, $y(k+1)$, would suffice), we could calculate the ideal control signal for the desired response.

Note that if the process has a delay of $d$ samples between the input and output, the model would be given by $y(k+1+d) \,=\, a y(k+d) + b u(k)$, such that the control law would be given by $u(k) \,=\, (1/b) y(k+1+d) - (a/b) y(k+d)$. In this second case, we would need to know the future values of the output $d$ and $d+1$ steps ahead of the current instant. However, both of these control laws are not feasible since the future values of the output are not known in advance.

Therefore, we move on to the second question raised earlier. How can we approximate the future behaviour of a variable over time? The simplest way to estimate the future value of a variable based on the available information at the current instant is to approximate the curve that describes the process variable's behaviour with a tangent line at that instant, and then consider the tangent line as an estimate for the future behaviour of the variable.

This procedure is analogous to the methodology employed in linearization. Therefore, \textcolor{red}{consider an arbitrary signal that varies over time, denoted as $x(t)$.} To estimate the value of $x(t+\delta t)$, where $\delta t$ is a time increment into the future, we can make the following approximation based on a first-order Taylor expansion:
\begin{eqnarray} \label{EqTaylorAcaoD}
x(t+\delta t) &\approx& x(t) + \frac{dx(t)}{dt} \delta t \, \text{.}
\end{eqnarray}

Using the approximation presented in Eq. \eqref{EqTaylorAcaoD}, we make an approximate prediction for the variable $x$ at the future time $t+\delta t$ based on the instantaneous value of the variable $x(t)$, which is known, and its time derivative evaluated at the instant $t$, which is also known. We can apply this approximation to the error signal of a control system. In this case, denoting $\hat{e}(t+\delta t)$ as the future estimate of the error, we have:
\begin{eqnarray} \label{EqTaylorAcaoD2}
\hat{e}(t+\delta t) &\approx& e(t) + \frac{de(t)}{dt} \delta t \, \text{.}
\end{eqnarray}

Therefore, using Eq. \eqref{EqTaylorAcaoD2} to estimate the error at a future time based on a linear combination of the error signal and its derivative (evaluated at the current time instant), we can consider a control action proportional to the future error signal of the system, at a time $T_d$ ahead of $t$, as follows:
\begin{eqnarray}\label{EqTaylorAcaoD3}
u_D(t) &=& K_D\hat{e}(t) \,\,=\,\, K_De(t) + K_D\frac{de(t)}{dt}T_d \, \text{.}
\end{eqnarray}

The control law presented in Eq. \eqref{EqTaylorAcaoD3} is of the proportional-derivative (PD) type, as it includes a term proportional to the error signal with gain $K_D$ and another term proportional to the derivative of the error signal with gain $K_D T_d$.

\begin{figure}[htb]
	\centering
	\includegraphics[width=0.6\linewidth]{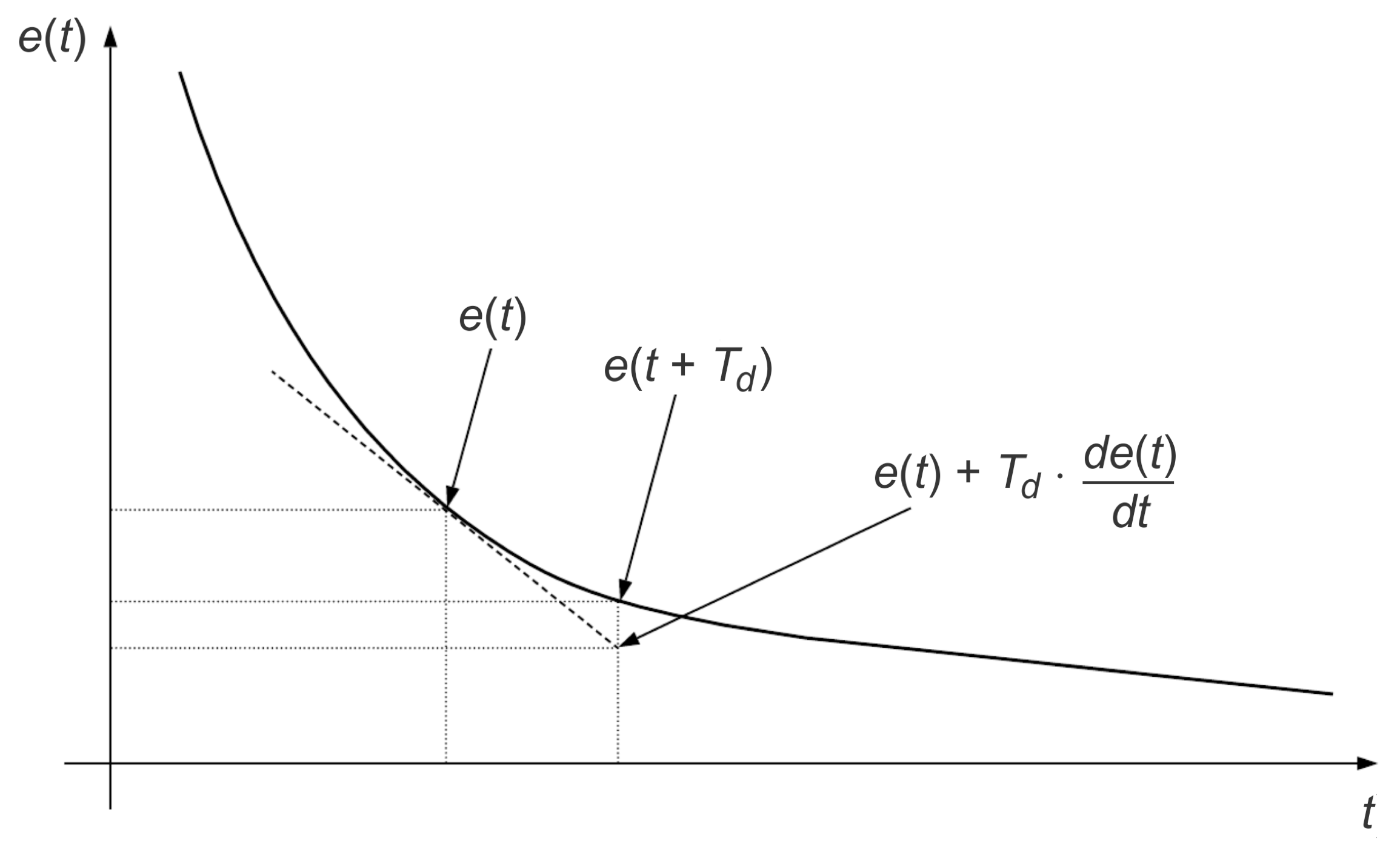}
	\caption{Derivative action as a linear prediction of the future error.}
	\label{acaoD}
      \end{figure}

As a complement to the previous discussion, Figure \ref{acaoD} graphically illustrates the evolution of an error signal over time $e(t)$ and the use of the approximation given in Eq. (2) to calculate $e(t+T_d)$ based on the tangent line at the point $t$. As we can observe in this figure, the signal $\hat{e}(t+T_d)$ reasonably approximates the real signal $e(t+T_d)$. However, it is evident that there is a difference between the actual and predicted values for the future error signal, depending on the shape of the curve $e(t)$ and the value of the prediction step $T_d$.

\FloatBarrier

Based on this idea, we can develop a controller that takes into account the present (proportional action - P), past (integral action - I), and future (proportional-derivative action - PD) actions in a unified manner, as follows:
\begin{eqnarray}\label{PID1}
u(t) &=& \overbrace{K_{p_1} e(t)}^{\text{P action}} + \overbrace{K_i \left(\int_{0}^t e(\mu) d\mu\right)}^{\text{I action}} + \overbrace{K_D e(t) + K_D \frac{de(t)}{dt} T_d}^{\text{PD action}} \, \text{.}
\end{eqnarray}

The equation \eqref{PID1} is rewritten as a PID control law, as follows:
\begin{eqnarray}\label{PID2}
u(t) &=& \overbrace{K_{p} e(t)}^{\text{P part}}+ \overbrace{K_i \left(\int_{0}^t e(\mu) d\mu\right)}^{\text{I part}} + \overbrace{K_d \frac{de(t)}{dt}}^{\text{D part}} \, \text{,}
\end{eqnarray}
\noindent for which we group the two proportional terms using $K_p\,=\,K_{p_1}+K_D$ and defining the derivative gain as $K_d\,=\,K_D T_d$.

\subsection{PID structures}

Based on the previous discussions, the instructor can demonstrate to students how the PID control law is equivalent to a PI control law with the addition of a derivative term, used as an approximation of the future value of the error signal.

\begin{figure}[htb]
	\centering
	\includegraphics[width=0.8\linewidth]{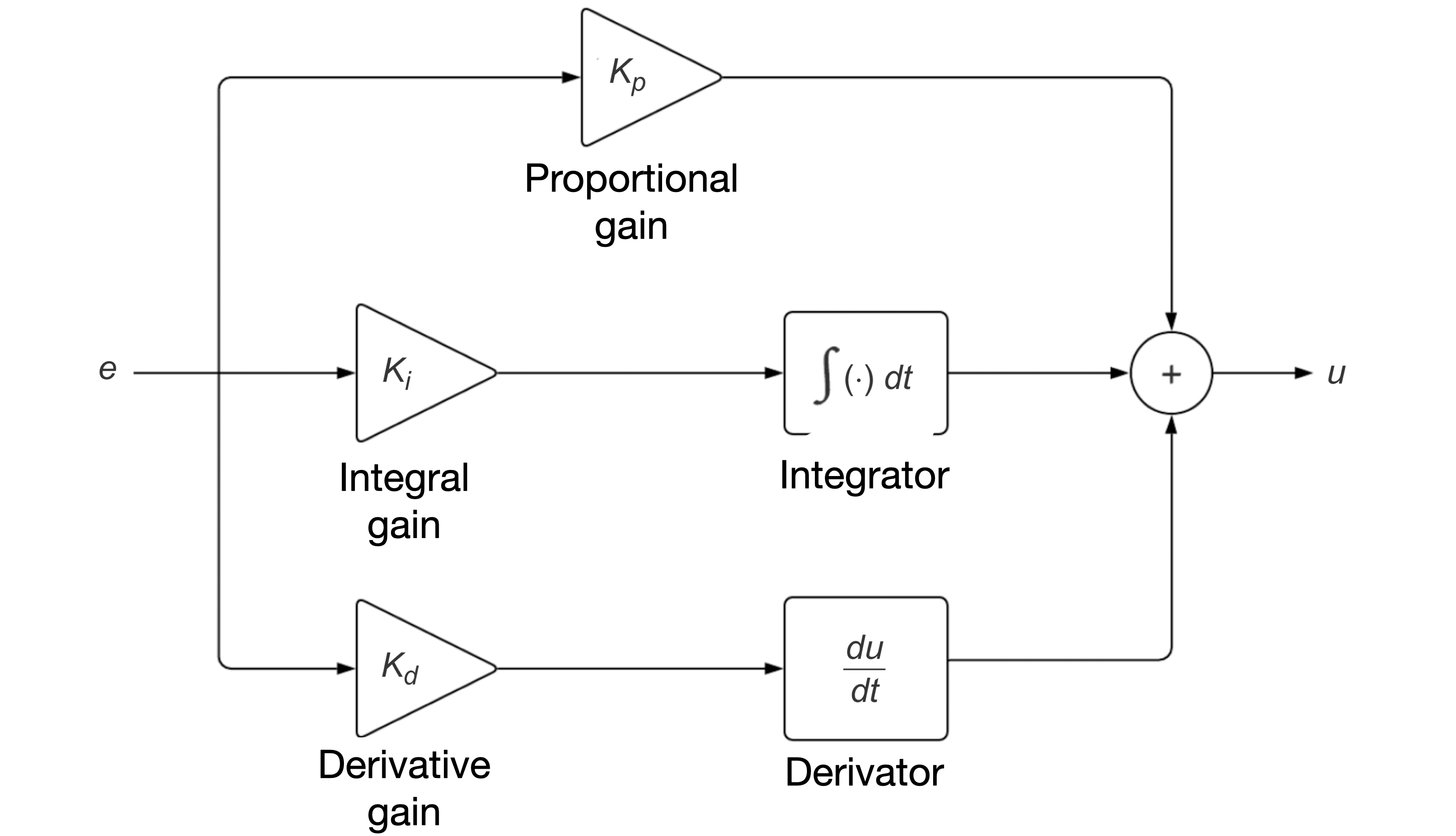}
	\caption{Ideal PID in a parallel setup.}
	\label{PIDparalelo}
\end{figure}

It should be noted that the PID configuration presented in Equation \eqref{PID2} is referred to as the \emph{ideal} PID and is implemented in the \emph{parallel} form. It is called ideal because we consider a perfect implementation of the derivative action in the control equation. We will discuss this fact in more detail shortly. Furthermore, this configuration is commonly referred to as parallel because the three control actions (proportional, integral, and derivative) are calculated separately and then summed together, as shown in the block diagram presented in Figure \ref{PIDparalelo}.

The ideal PID control can also be implemented in other forms. In our educational proposition, we mention another one known as the \emph{academic} structure. This implementation is commonly used in most control systems textbooks and teaching materials. The expression for the academic PID control law is given by:
\begin{eqnarray}\label{PID3}
u(t) &=& K_{c} \left( e(t) + \frac{1}{T_i} \left(\int_{0}^t e(\mu) d\mu\right) + T_d \frac{de(t)}{dt} \right) \text{.}
\end{eqnarray}

\begin{figure}[htb]
	\centering
	\includegraphics[width=0.8\linewidth]{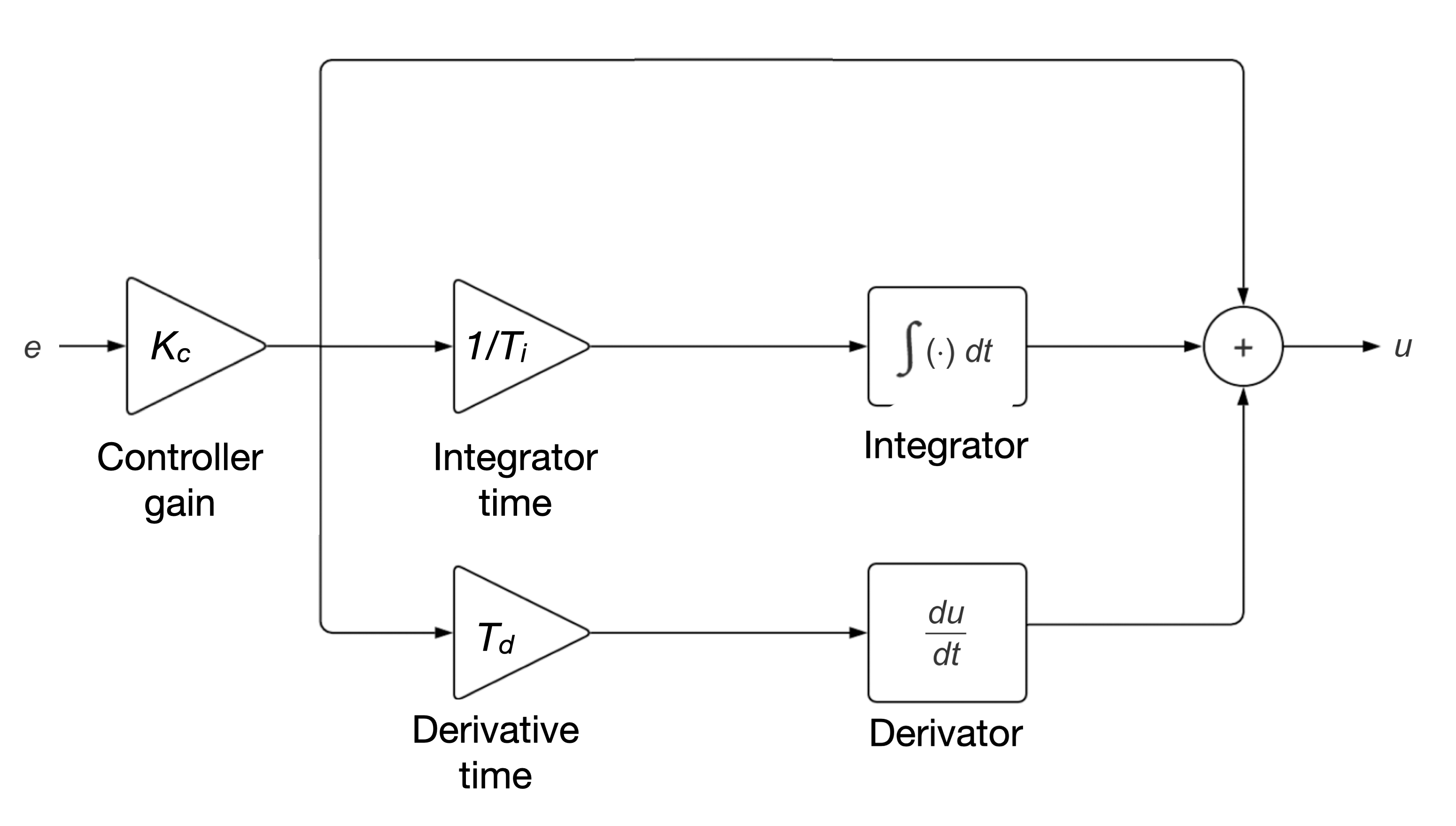}
	\caption{Ideal PID in an academic configuration.}
	\label{PIDacademico}
\end{figure}

At this point, it is worth noting that this expression is similar to the PI control law with the addition of the derivative action. In this case, we have a weighting factor $T_d$ associated with the derivative action, which is called the \textbf{derivative time} parameter. Figure \ref{PIDacademico} illustrates the implementation of this control structure (in the ideal case). It should be emphasized that the gain $K_c$ multiplies all terms in this control law. Note that the integral is defined from $\mu = -\infty$ to $\mu = t$. However, since we only consider causal signals defined from $t\,=\,0$, we obtain $\int_{-\infty}^te(\mu)d\mu) \,=\,\int_{0}^te(\mu)d\mu)$, given that $\int_{-\infty}^0e(\mu)d\mu) \,=\, 0$.

In both of these PID controller configurations, we have three tuning parameters: the gains $K_p, K_i$ and $K_d$ in the first case, and the gain $K_c$ and the times $T_i$ and $T_d$ in the second case. Similar to the tuning of PI controllers, the teaching approach described here involves presenting and discussing different PID tuning methodologies from the literature, such as IMC and Ziegler-Nichols, without addressing, for example, analyses using root locus techniques.

Another particularly important practical aspect that is introduced to undergraduates, when discussing PID controllers, is that the implementation of the derivative action can be set only with respect to the process variable. 

It is simple to demonstrate that standard PID structures (as the ones shown in the prequel) tend to generate very large (theoretically infinite) control action when the reference signal changes abruptly (with a step-like behaviour, for instance). Given the fact that such large variation of the control signal is not interesting in any real application, the derivative action, at first given by $K_d\frac{de(t)}{dt}$, can be simply replaced by $-K_d\frac{dy(t)}{dt}$ in order to avoid the variations implied by $\frac{dr(t)}{dt}$. 

With regard to this matter, it is also important to elaborate on the fact that, when considering the objective of disturbance rejection, this simplification of the derivative action does not compromise the obtained responses, given that the rejection response is based on the process output variations only (and not on the reference signal). Based on these discussions, undergraduates can understand why the derivative action is only set over the output in practical applications.

\subsection{Noise filtering}

Another important topic to be discussed is the noise observed when measuring the process variable. The student should analyse practical experiments and understand that, due to the peculiarities of the process (e.g., turbulence in a fluid) or measurement systems (electromagnetic interference), the signals received by the controller exhibit rapid variations and low-amplitude fluctuations superimposed on the desired value being measured.

Once these signals are characterised, the effect they have on the control signal should be observed. In the case of PI control, it can be easily explained that only the P action will amplify the noise, as the integral calculates an average value of the noise, which is small and therefore negligible.

Amplification of noise can have negative effects, as the actuator will have to respond to rapid and large-amplitude signals. As a result, the actuator may be damaged. To reduce this problem, there are two alternatives: either use a smaller gain $K_p$, which affects the control performance, or attenuate the noise signal in the measured signal.

For the second option, the idea of filtering the measured signal can be introduced to separate the desired measurement from the undesired variations. The simplest way to present the topic of filtering is through experiments that demonstrate that the average of the noisy signal varies much less than the instantaneous value. Thus, a simple digital moving average filter (with a window of size $M$) can be easily implemented and adjusted. Empirical adjustments of the window size $M$ can be used to show the trade-off between noise attenuation and deterioration of measurement quality.

With the introduced concepts, the student will be able to understand and implement digital control strategies for simple processes, considering the most important aspects encountered in practice, \textcolor{red}{given that on-off and PI/PID controllers are undoubtedly the most used in real applications.}

\section{An experimental test-bench}
\label{sec8}

\noindent In addition to the discussed educational approach, our teaching proposal also aims for integration with practice. Thus, in the courses taught based on the detailed methodology, students use a low-cost experimental kit to validate the introduced concepts.

\begin{figure}[htb]
    \centering
    \includegraphics[width=\linewidth]{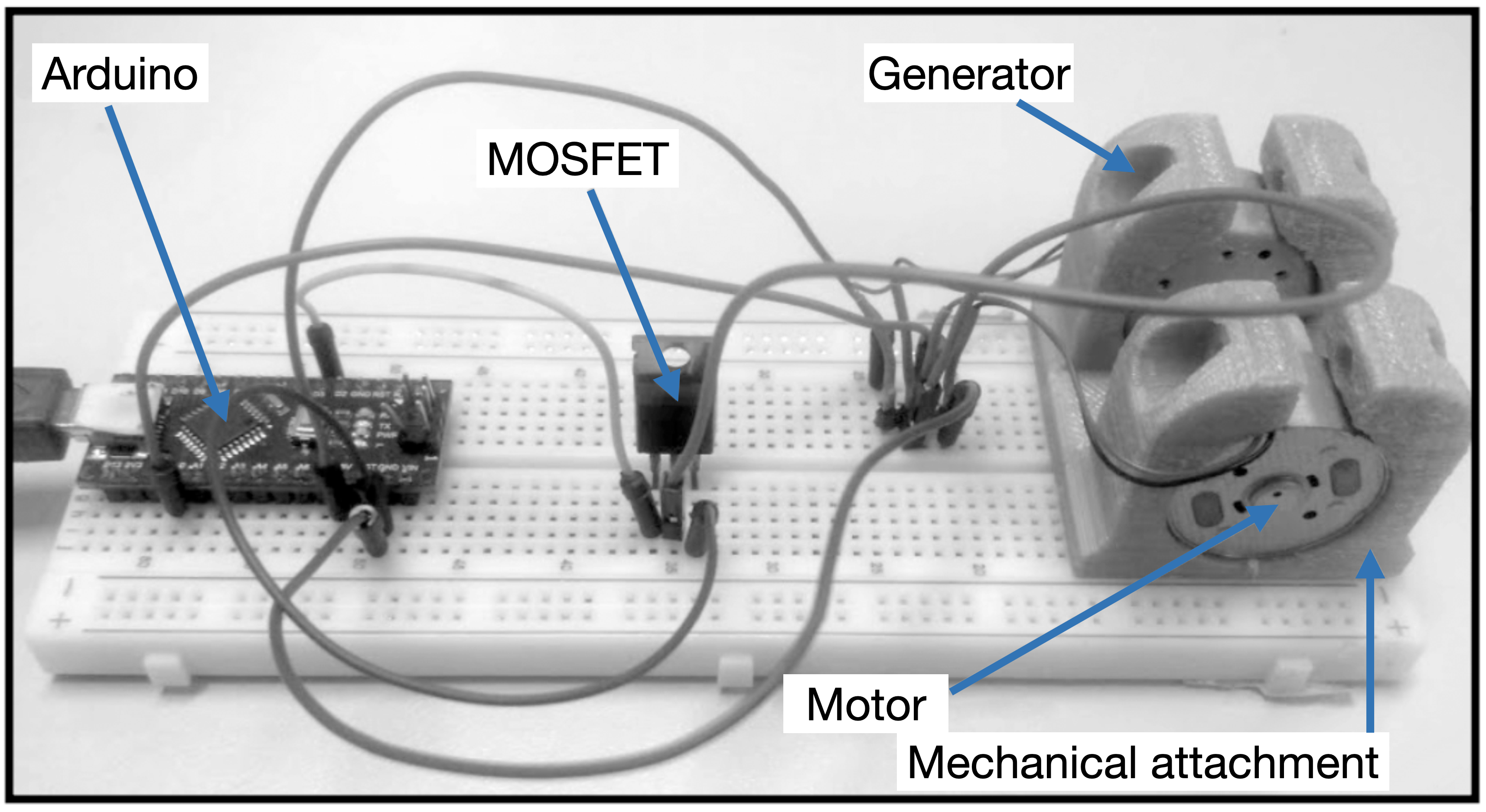}
    \caption{Photo of an experimental kit.}
    \label{fotoKIT}
\end{figure}

In Figure \ref{fotoKIT}, we present such a kit, composed of:
\begin{itemize}
    \item two $5$V DC motors;
    \item a MOSFET transistor;
    \item a small breadboard;
    \item copper wires (jumpers);
    \item an Arduino micro-controller.
\end{itemize}

In this kit, the two motors are mechanically coupled through their shafts. So, when one motor is activated by a voltage $V_i(t)$ sent by the Arduino (control signal), the other motor moves, generating a voltage $V_o(t)$ across its terminals (output signal). The activation of the first motor is done through the PWM port of the Arduino, which drives the MOSFET as a switch, in such a way that this motor is subjected to an average voltage value between $0$ and $5$ V. On the other hand, the output $V_o(t)$ at the terminals of the second motor is directly connected to an \textcolor{red}{analog} port of the Arduino.

\begin{figure}[htb]
    \centering
    \includegraphics[width=\linewidth]{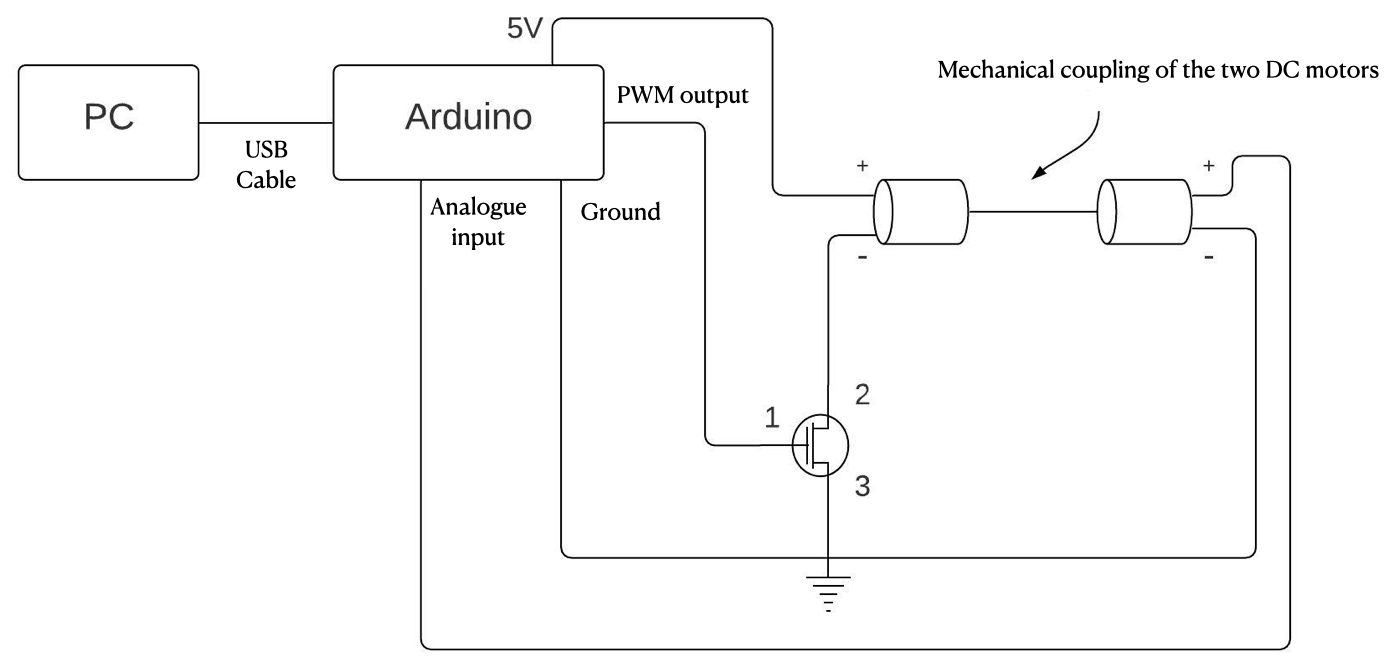}
    \caption{Electrical-mechanical implementation diagram of the experimental test-bench.}
    \label{circuitoKit}
\end{figure}

\textcolor{red}{
\begin{remark}
Regarding implementation aspects of the proposed experimental test-bench, Figure \ref{circuitoKit} provides an electrical-mechanical diagram of the proposed kit.    
\end{remark}
}

All the control strategies analysed can be implemented on the Arduino micro-controller, which already has an \textcolor{red}{analog}-to-digital conversion system and allows communication with the computer. On the computer, the control codes can be written in a simple way, and graphs of the variables of interest can be viewed.

Some advantages of this kit that can be highlighted from the perspective of illustrating concepts are: (i) it allows to view of the rotational speed of the motor shafts and practical disturbances by manually altering the motor shafts; (ii) the system has a nonlinear static characteristic and the output signal is noisy; (iii) the process dynamics are fast, allowing for multiple experiments in a short period of time; (iv) in the central operating region, a first-order model represents the system behaviour well; (v) all the control strategies analysed in our educational approach can be taken into account, in addition to the practical verification of their properties.

\section{A simulation software}
\label{sec8b}

\noindent Complementary, we also propose an open-source simulation software that can be used to validate theoretical concepts. Specifically, we stress that the use of interactive tools has become increasingly common, especially for educational purposes, as they allow for immediate visual representation of the process's evolution based on the given commands, bringing new possibilities to be explored \citep{refinte}.

The popularisation of interactive tools is due to the considerable processing power of personal computers, which allows for the execution of heavier software without the need for a server or cluster for processing. Another reason for their popularity is the availability of free packages for creating programs and graphical interfaces. Here, we present a set of interactive simulation tools developed using \emph{Tkinter}, \emph{Matplotlib}, and \emph{Numpy} packages available for \emph{Python} \citep{matplotenumpybook}.

Accordingly, this Section describes the set of interactive tools developed for our educational approach. \textcolor{red}{The detailed tools are open source, and available for download along with source code (refer to \emph{Code availability} declaration, at the end of this article)}. These tools can be used in control courses in order to discuss the concepts associated with basic control systems, implementation details of the considered controllers, and analysis of techniques to correct undesired behaviours of each type of control scheme (in this case, on-off, P, PI and PID controllers). Note that, for simplicity and better understanding for the part of students, one specific environment was developed for each kind control strategy, in such a way that the undergraduate can focus on the particular properties of each controllers.

\begin{figure}
    \centering
    \includegraphics[width=\linewidth]{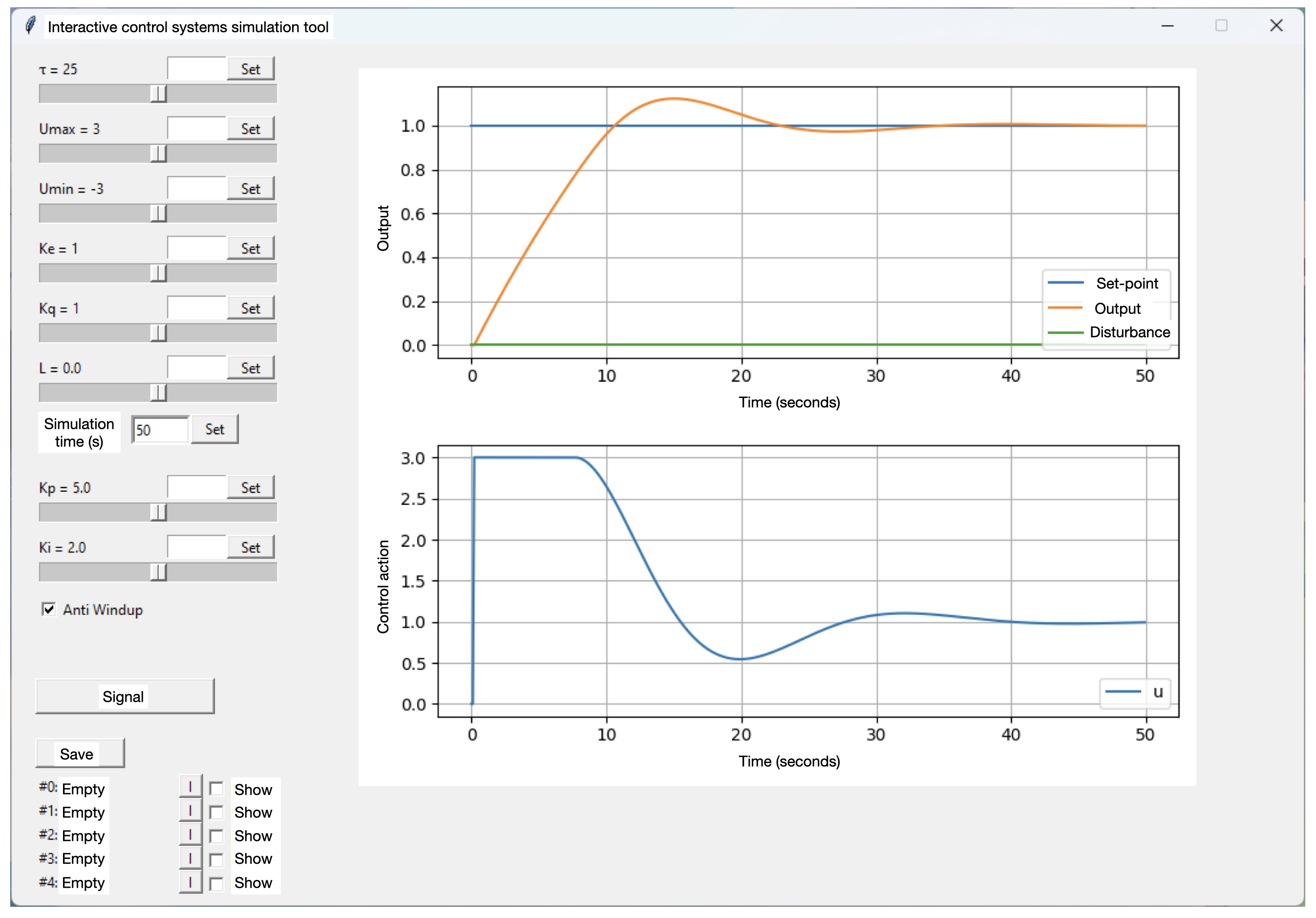}
    \caption{Graphical interface of the interactive tool: PI controller with an anti-windup strategy.}
    \label{fig:interfaceawp}
\end{figure}

The graphical interface aims to allow \textcolor{red}{the student} to vary the parameters of the controllers and interactively view the impact that these changes have on the process variables.
In a generic way, the graphical interface can be divided into blocks that will be discussed below and can be seen in Figure \ref{fig:interfaceawp}; in this case, we present a PI controller with anti-windup.

On the right side of the main window, two graphs are displayed: the first one shows the evolution of the process variable, reference, and disturbance applied to the process, and the second one shows the control signal (see Figure \ref{fig:interfaceawp}).

In this part of the graphical interface, in the left frame of the main window, the objective is to allow the user to vary the parameters of the model and the control in real-time and observe the immediate impact of these changes. To achieve this, the tool provides sliders that enable the user to make these adjustments, as shown in Figure \ref{fig:interfaceawp}.

The parameters that can be adjusted may vary from tool to tool, depending on the controller being used. However, for all controllers, there is the adjustment of the plant model parameters, which include: $\tau$, the time constant in seconds; $u_{\text{max}}$, the upper saturation limit of the controller; $u_{\text{min}}$, the lower saturation limit of the controller; $K_e$, the static gain related to the set-point; $K_q$, the static gain related to the disturbance, and $L$, the process delay in seconds. In addition to these adjustments, there is a time scale adjustment, where the user can input the maximum simulation time in seconds. It is important to note that changing the time scale will delete all saved wave-forms. The remaining parameters vary from controller to controller, but they all follow the same logic: there is a slider for each controller parameter. For example, in Figure \ref{fig:interfaceawp}, as it is a simple PI control, we only have the adjustment of the proportional and integral gains.

In the lower left corner of the main window, there is a menu for signal creation. By clicking the ``\emph{Signal}'' button, a new window will appear where the user can define a waveform of interest from the available options. There are check-boxes that allow using the created signal as the set-point (``Use as set-point'' checkbox), disturbance signal (``Use as disturbance'' checkbox), or control signal (``Use as control signal'' checkbox). It is worth noting that by checking the last checkbox, the tool assumes an open-loop behaviour, meaning that the controller action is decoupled from the process model dynamics.

In this window, it is also possible to add noise to the simulation by checking the ``Add noise'' checkbox. The existence of noise can be configured starting from a specific time $t_0$ and with an amplitude defined by the ``Amp\_n'' slider.

\begin{figure}
    \centering
    \includegraphics[width=\linewidth]{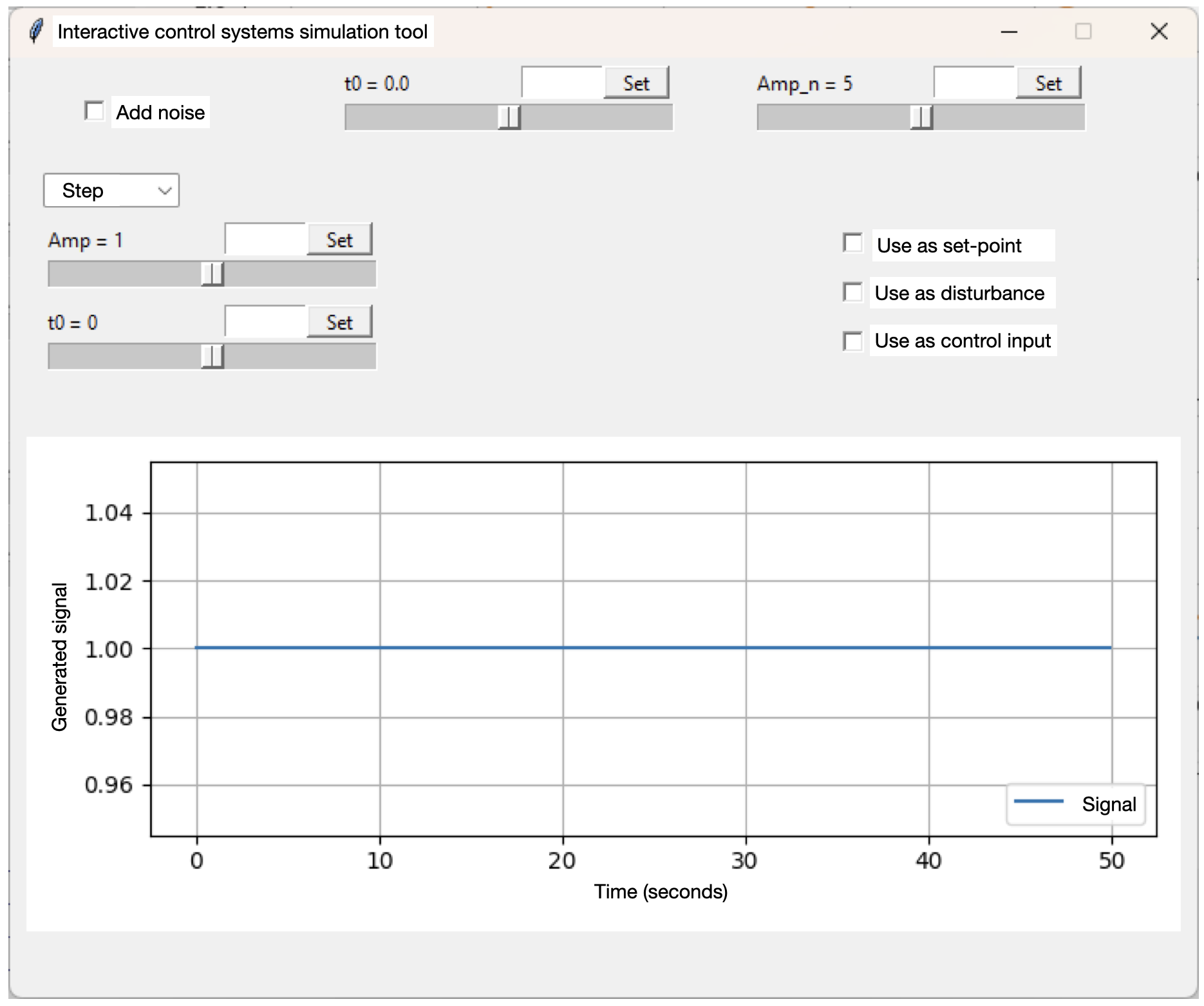}
    \caption{Window after hitting the \emph{Signal} button.}
    \label{fig:sinal}
\end{figure}

Several waveform options are available for simulation, including: step, sine, triangular, or square wave. For each waveform, sliders are provided to adjust specific parameters. This can be observed in Figure \ref{fig:sinal}, where we show the selection of a step signal.

\begin{figure}
    \centering
    \includegraphics[width=0.5\linewidth]{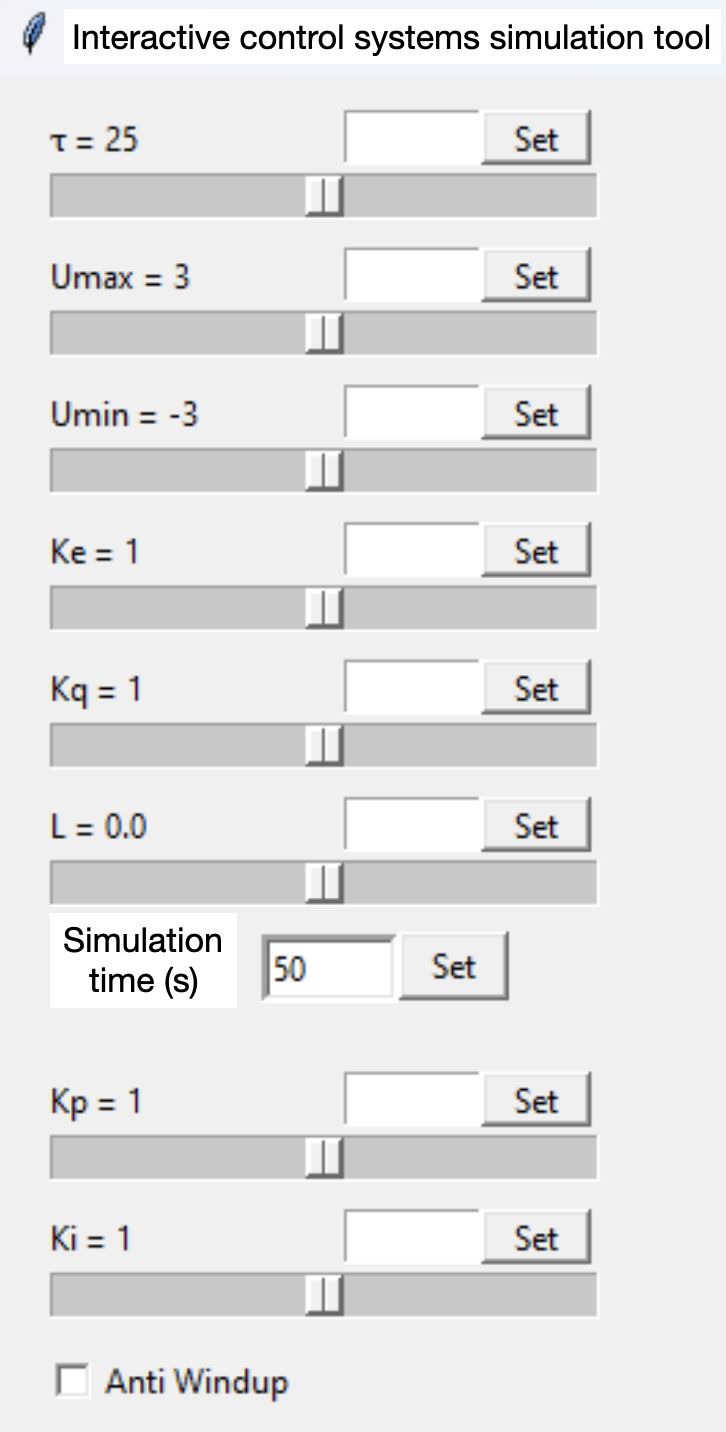}
    \caption{Model and controller parameter sliders in the interactive simulator.}
    \label{fig:paramsawp}
\end{figure}
    
In the last section at the bottom left of the main window, there is a menu for saving the results of an experiment in order to make comparisons.

By clicking the ``\emph{Save}'' button and choosing one of the available options to display, the tool retains the results of the performed simulation, which can be compared with subsequent simulations. The user can give a name to each simulation they want to save and choose which one to display.

\begin{figure}[h]
    \centering
    \includegraphics[width=0.4\linewidth]{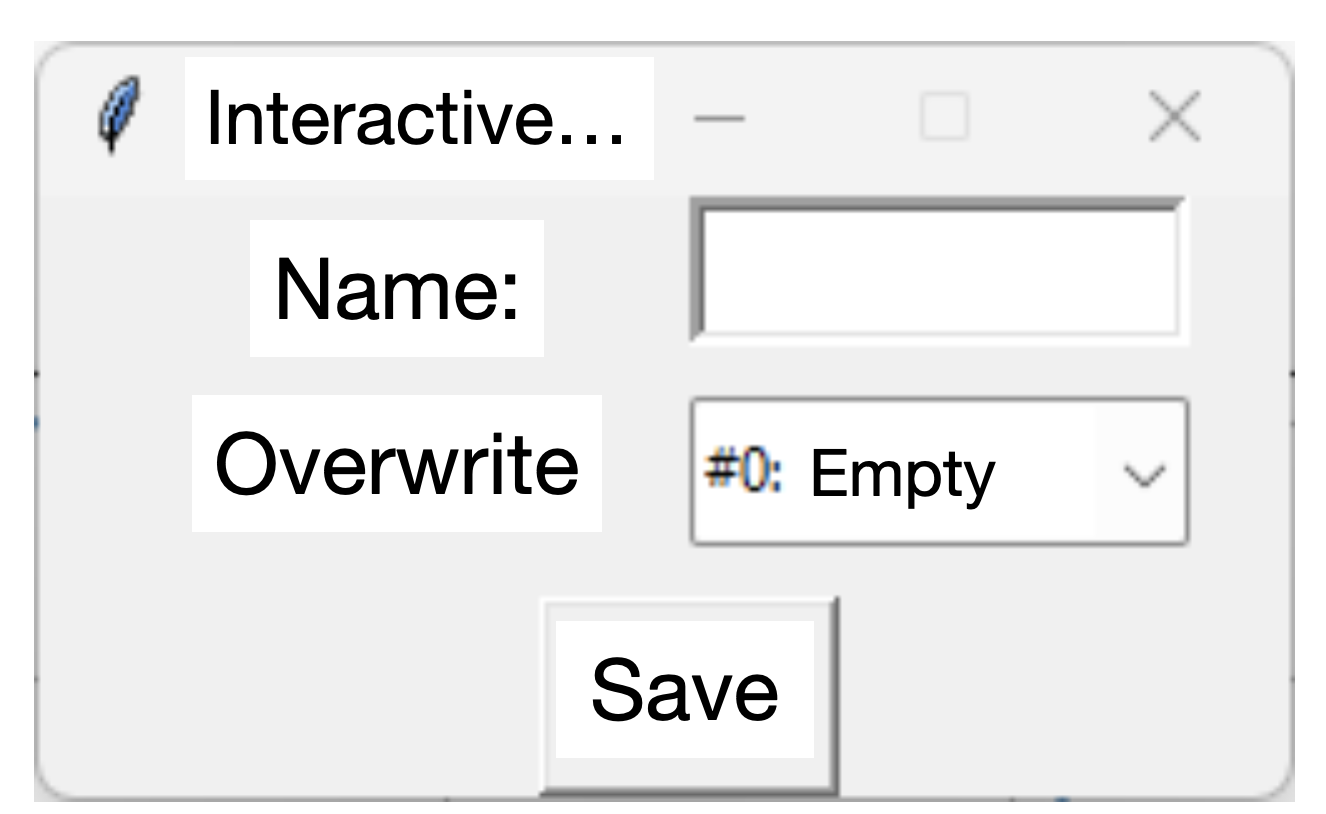}
    \caption{Window used to save wave-forms.}
    \label{fig:salvarfig}
\end{figure}

\subsection{An application example}

To demonstrate the use of the tool, this section presents an example comparing the behaviour of a PI controller with and without \emph{anti-windup} action. It is assumed that the PI control has already been studied, and its implementation and adjustment have been analysed with the corresponding tool. The tool for PI with \emph{anti-windup} is specific to the understanding and use of the windup phenomenon and the \emph{anti-windup} strategy. The PI controller uses the following settings: $K_p = 5$, $K_i = 4$, considering the process parameters as $\tau = 15$ s, $u_{\text{max}} = 3$, $u_{\text{min}} = -3$, $K_e = 1$, $K_q = 1$, and $L = 0$.

In this example, the student should adjust the values of the plant and control parameters in the main window using the corresponding sliders. After that, they should define the simulation scenario by selecting the signals applied to the process, which in this case are a unit step in the reference and no disturbances.

\begin{figure}
    \centering
    \includegraphics[width=\linewidth]{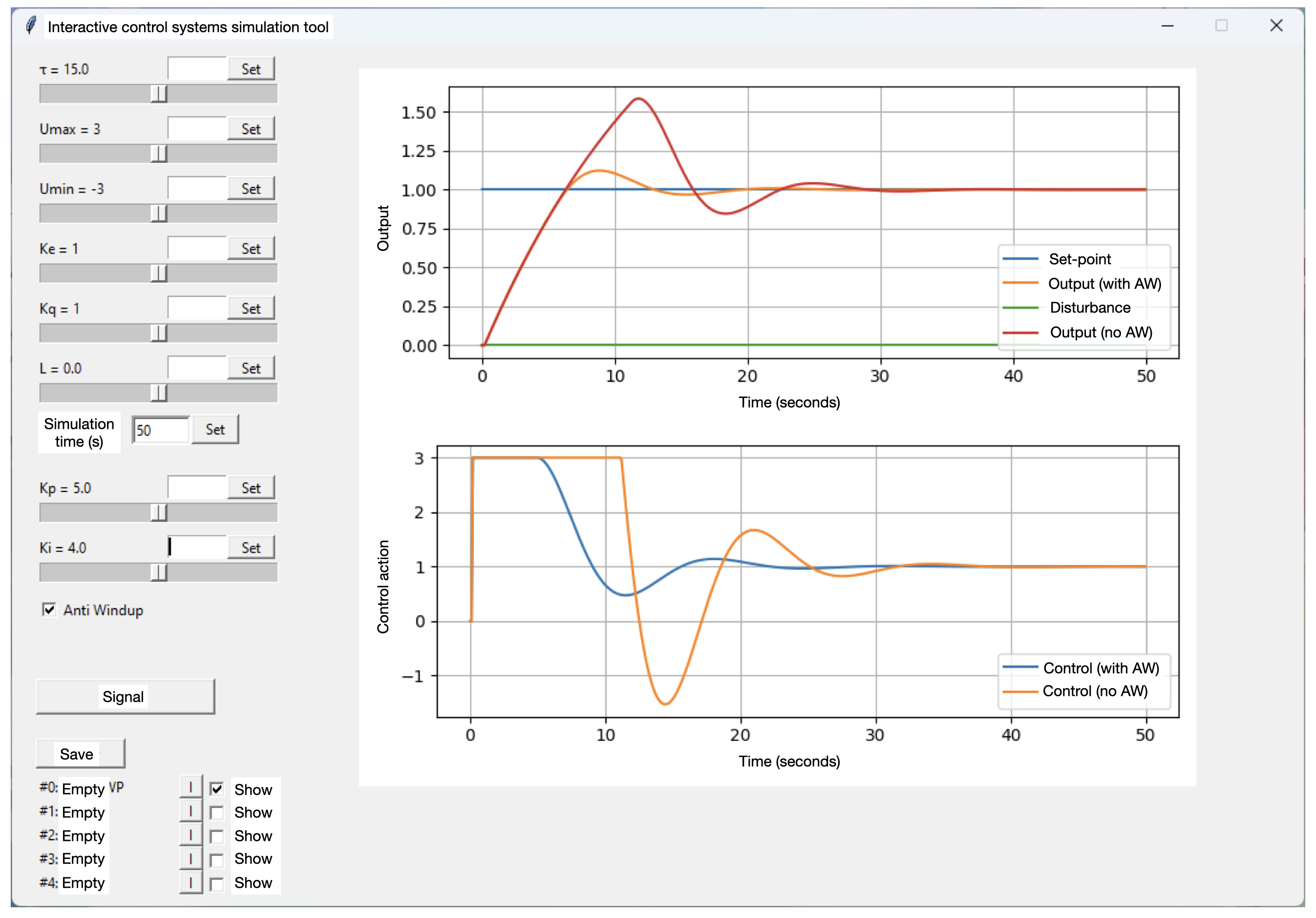}
    \caption{Comparison of the example system: Process output and control signal with \emph{anti-windup} (orange) and without (red).}
    \label{fig:semantiwindup}
\end{figure}

In the first part of the experiment, the student observes the effect of windup, noting that even after the system output exceeds the set-point, the control action remains saturated in the PI controller without Anti-Windup action. This leads to a response of the process variable with a high overshoot. Additionally, a long settling time is observed in this situation. With the tool, the student can analyse the effect of tuning the PI controller on this response. For example, they can adjust the system for a slower response with less overshoot. \textcolor{red}{The student} can also analyse the effect of the value of $u_{\text{max}}$ (control signal upper bound) on the system's response.

For comparison purposes, the second part of the experiment involves simulating the system with the Anti-Windup action enabled and comparing it with the previous result. Therefore, the \textcolor{red}{student} should save the already obtained result using the save menu, as explained earlier, before activating the new control with \emph{anti-windup}.

As shown in Figure \ref{fig:semantiwindup}, in this case, the overshoot is considerably smaller compared to the previous controller, as the actuator exits saturation more quickly since the control action does not accumulate \textcolor{red}{integral} action after the controller saturation.

\section{Teaching-learning results}
\label{sec9}

\noindent Taking into account the previous details on the syllabus and teaching framework of the proposed \emph{Introduction to process control} course, we discuss, next, some results obtained from the teaching-learning process. 

We evaluate these results, specifically, considering the responses provided by undergraduate students under randomised anonymous interviews\footnote{In particular, sixty nine undergraduate students took part in these interviews. Each one of the following tables presents a question that was asked and the percentage of students that marked each possible response. }. We also base our assessments on the teachers' perception of the progress of the proposed activities in classroom, and on testimonials from the undergraduate students regarding motivation and performance. Without claiming to be conclusive, we elaborate some reflections based on experiences from the the period that covers $2016$ up to $2023$.

The proposed educational approach was first included to the CAE curriculum from UFSC in $2016$, after an undergraduate curricular reform; the \emph{Introduction to process control} course has been positioned, since then, in the third semester of the \textcolor{red}{program} (out of ten), followed by other control courses along the following terms (refer to the full curriculum as presented in Appendix \ref{appendiceACursos}). At first, the course had an weekly classroom load of two hours, with practical activities and experiences to be done by the students as extra assignments (considering the modular test kit detailed in Section \ref{sec8}). Later, from $2018$ onward, the course's workload was increased, including an additional hour of laboratory activities per week, which was focused on monitoring the \textcolor{red}{students'} progression on the experimental activities.

As elaborated in the introduction, the expected results from the proposed educational approach was, essentially, to motivate the undergraduate students on the many topic of control systems and, at the same time, make practical and application-oriented use of basic notions of maths and physics, which are usually taught merely \textcolor{red}{as} tool-utensils, often disconnected from reality and practical applications.

Furthermore, by including the proposed course to our CAE undergraduate curriculum at UFSC, we expected to see students progressing along control-related subjects (such as: \emph{Signals and linear systems}, from the fourth semester, \emph{Modelling and simulation of processes}, from the fifth semester, and  \emph{Control systems}, from the sixth semester) with a broader and clearer perspective on the importance of control systems and, also, with greater motivation to study the related more \textcolor{red}{advanced} mathematical tools, such as Laplace and $\emph{Z}$ transforms and so forth.

Finally, we also expected to see a better approval rate in the \emph{Control systems} course, from the sixth semester, which had been exhibiting an excessively high rate of failure.

\noindent\textbf{Regarding motivation}: \\
From the conducted interviews and testimonials, we can conclude that the \emph{Introduction to process control} course is, indeed, a successful educational approach to motivate undergraduate students in topics of control systems. Over $89\,\%$ of interviews students indicate that the course fulfils, at least in parts, the objective of introduction, in the beginning of the \textcolor{red}{\textcolor{red}{undergraduate program}}, the basic notions of control (see Table \ref{questionariop1}); furthermore, over $97\,\%$ of them think that it a course of this nature is important for their undergraduate studies (see Table \ref{questionariop2}). 

\begin{table}[ht]
    \centering
      \caption{How much do you think that \emph{Introduction to process control} course fulfils the objective of introducing, in the beginning of the \textcolor{red}{program}, the basic concepts of control systems?}
\label{questionariop1}
    \small
    \begin{tabular}{|p{1.05in}|p{0.65in}|p{0.65in}|p{0.65in}|p{0.65in}|}\hline
        \hline Number of answers & A lot & Reasonably & Little & Nothing \\  \hline
       $69$ & $53.6\,\%$ & $36.2\,\%$ & $10.1\,\%$ & $0\,\%$ \\ 
         \hline \hline
    \end{tabular}
\end{table}

\begin{table}[ht]
    \centering
      \caption{How important do you think it is to have a course of this nature, focused on control systems, already in the beginning of the \textcolor{red}{\textcolor{red}{undergraduate program}}?}
\label{questionariop2}
    \small
    \begin{tabular}{|p{1.05in}|p{0.65in}|p{0.65in}|p{0.65in}|p{0.65in}|}\hline
        \hline Number of answers & A lot & Reasonably & Little & Nothing \\  \hline
       $69$ & $85.5\,\%$ & $11.6\,\%$ & $2.9\,\%$ & $0\,\%$ \\ 
         \hline \hline
    \end{tabular}
\end{table}

With respect to this matter, we note that, from the professor perspective, a great interest from the part of the students was systematically (and evidently) observed in topics related to control applications in the real world, as well as in practical activities. We emphasise that the proposed educational approach consists, during classes, in illustrating the majority of the control-related concepts with the aid of real industrial problems, brought from project experiences with companies or laboratory research and development.
 
 Nevertheless, we also stress that we could observe a systematic tendency of students to lose interest when formalising ideas and concepts with mathematical tools. In these cases, the portion of students that remained avidly motivated and interested was always smaller. In Table \ref{questionariop13}, the interviewed students indeed rank these (mathematically tougher) subjects as the harder form them to grasp.

\textcolor{red}{\begin{table}[ht]
    \centering
      \caption{What content of the \emph{Introduction to process control} course did you have the most difficulty with?}
\label{questionariop13}
    \small
    \begin{tabular}{|c|c|}
    \hline
        \hline Number of answers & $67$ \\ \hline Maths in general & $22.4\,\%$ \\ \hline Control concepts &$13.4\,\%$\\ \hline ODEs and DEs & $20.9\,\%$ \\ \hline Models and their physical relations & $13.4\,\%$ \\ \hline Static processes characteristics &$1.5\,\%$ \\ \hline Dynamic processes characteristics & $19.4\,\%$ \\ \hline
      PI control & $9\,\%$ \\ 
         \hline \hline
    \end{tabular}
\end{table}}

 However, we note that the introduction of a control course in the early terms of the CAE degree instigated several students to move towards control-related topics in research initiation or internship activities (in recent years, most students have been to work in other areas, such as programming and informatics). Indeed, even though many of the interviewed undergraduates (over $60\,\%$) evaluate the \emph{Introduction to process control} syllabus to be somewhat complex (see Table \ref{questionariop7}), the vast majority of them (over $79\,\%$) think that the course helped them to understand the relevance of studying physics and mathematical tools in engineering (see Table \ref{questionariop3}); in some level, the \emph{Introduction to process control} course also motivated (over $62\,\%$ of) students to reinforce their knowledge on maths and physics (see Table \ref{questionariop4}).

 \begin{table}[ht]
    \centering
      \caption{Regarding the content/syllabus of the \emph{Introduction to process control} course, do you consider it to be:}
\label{questionariop7}
    \small
    \begin{tabular}{|p{1.05in}|p{0.65in}|p{0.65in}|p{0.65in}|p{0.65in}|}\hline
        \hline Number of answers & Very complex & Complex & Adequate & Very basic \\  \hline
       $69$ & $26.1\,\%$ & $34.8\,\%$ & $39.1\,\%$ & $0\,\%$ \\ 
         \hline \hline
    \end{tabular}
\end{table}

\begin{table}[ht]
    \centering
      \caption{How much do you consider that the \emph{Introduction to process control} helps to understand the importance of the mathematics and physics content taught in high school and in the first semesters of the \textcolor{red}{program}?}
\label{questionariop3}
    \small
    \begin{tabular}{|p{1.05in}|p{0.65in}|p{0.65in}|p{0.65in}|p{0.65in}|}\hline
        \hline Number of answers & A lot & Reasonably & Little & Nothing \\  \hline
       $69$ & $37.7\,\%$ & $42\,\%$ & $15.9\,\%$ & $4.3\,\%$ \\ 
         \hline \hline
    \end{tabular}
\end{table}

\begin{table}[ht]
    \centering
      \caption{After taking the \emph{Introduction to process control} course (or while taking it), how much more motivated did you feel to reinforce your knowledge of mathematics and physics taught in high school and in the first semesters of the \textcolor{red}{program}?}
\label{questionariop4}
    \small
    \begin{tabular}{|p{1.05in}|p{0.65in}|p{0.65in}|p{0.65in}|p{0.65in}|}\hline
        \hline Number of answers & A lot & Reasonably & Little & Nothing \\  \hline
       $69$ & $30.4\,\%$ & $31.9\,\%$ & $30.4\,\%$ & $7.2\,\%$ \\ 
         \hline \hline
    \end{tabular}
\end{table}

 Regarding the laboratory tasks and experimental essays, we highlight that the majority of undergraduates evaluate them to be adequate (over $68\,\%$ of students) and important (over $72\,\%$ of students), at some level, for their learning process (see Tables \ref{questionariop8} and \ref{questionariop9}, respectively), which indicates the necessity of practical counter-parts and validation experiences in control subjects as the proposed one. We note that the majority of students (over $60\,\%$) evaluate that the \emph{Introduction to process control} course has a fair theoretical-practical integration of contents (see Table \ref{questionariop10}).

\begin{table}[ht]
    \centering
      \caption{Regarding the laboratory activities of the \emph{Introduction to process control} course, such as numerical simulation exercises and experimental validations, do you consider that the activities are:}
\label{questionariop8}
    \small
    \begin{tabular}{|p{1.05in}|p{0.65in}|p{0.65in}|p{0.65in}|p{0.65in}|}\hline
        \hline Number of answers & Very complex & Complex & Adequate & Very basic \\  \hline
       $69$ & $2.9\,\%$ & $7.2\,\%$ & $68.1\,\%$ & $21.7\,\%$ \\ 
         \hline \hline
    \end{tabular}
\end{table}

\begin{table}[ht]
    \centering
      \caption{Regarding the importance of practical activities in simulation and experimental for your learning process along the \emph{Introduction to process control} course, you evaluate that they were:}
\label{questionariop9}
    \small
    \begin{tabular}{|p{1.05in}|p{0.65in}|p{0.65in}|p{0.65in}|p{0.65in}|}\hline
        \hline Number of answers & Very important & Important & Of little importance & Irrelevant \\  \hline
       $69$ & $20.3\,\%$ & $52.2\,\%$ & $21.7\,\%$ & $5.8\,\%$ \\ 
         \hline \hline
    \end{tabular}
\end{table}

\begin{table}[ht]
    \centering
      \caption{Regarding theoretical-practical integration of the \emph{Introduction to process control} course, you evaluate that it was:}
\label{questionariop10}
    \small
    \begin{tabular}{|p{1.05in}|p{0.52in}|p{0.52in}|p{0.52in}|p{0.52in}|p{0.52in}|}\hline
        \hline Number of answers & Very good & Fair & Merely sufficient & Weak & Very weak \\  \hline
       $69$ & $5.8\,\%$ & $36.2\,\%$ & $24.6\,\%$ & $29\,\%$ & $4.3\,\%$ \\ 
         \hline \hline
    \end{tabular}
\end{table}

\noindent\textbf{Regarding performance in other control-related courses}:\\ Bearing in mind the motivation of students with the \emph{Introduction to process control} course, we now assess their performances in other control-related courses along the CAE \textcolor{red}{\textcolor{red}{undergraduate program}}. First, we highlight that over $78\,\%$ of interviewed students indicate that this course was an important prerequisite, at some level, for the following courses in their \textcolor{red}{undergraduate program} (see Table \ref{questionariop5}). Furthermore, over $68\,\%$ of them evaluated that the course served to motivate for the following terms (see Table \ref{questionariop6}).

\begin{table}[ht]
    \centering
      \caption{How much do you consider that the \emph{Introduction to process control} course helps as a prerequisite for other the control-related courses, over the following terms of the \textcolor{red}{program}?}
\label{questionariop5}
    \small
    \begin{tabular}{|p{1.05in}|p{0.65in}|p{0.65in}|p{0.65in}|p{0.65in}|}\hline
        \hline Number of answers & A lot & Reasonably & Little & Nothing \\  \hline
       $64$ & $43.8\,\%$ & $34.4\,\%$ & $18.8\,\%$ & $3.1\,\%$ \\ 
         \hline \hline
    \end{tabular}
\end{table}

\begin{table}[ht]
    \centering
      \caption{How much do you consider that the \emph{Introduction to process control} course was motivating for the other the control-related courses, over the following terms of the \textcolor{red}{program}?}
\label{questionariop6}
    \small
    \begin{tabular}{|p{1.05in}|p{0.65in}|p{0.65in}|p{0.65in}|p{0.65in}|}\hline
        \hline Number of answers & A lot & Reasonably & Little & Nothing \\  \hline
       $64$ & $37.5\,\%$ & $31.3\,\%$ & $18.8\,\%$ & $12.5\,\%$ \\ 
         \hline \hline
    \end{tabular}
\end{table}

Considering the CAE curriculum at UFSC (refer to Appendix \ref{appendiceACursos}),  $94\,\%$ of interviewed students that took \emph{Signals and linear systems} had success in their first try, with an average semester grade of $6,3$ (out of $10$). Likewise, $88\,\%$ of undergraduates that took \emph{Modelling and simulation of processes} were successful in their first try. In terms of the students that took \emph{Control systems}, which had been facing a high failure rate, as previously stated, $70\,\%$ were successful in their first try. The average rate of approval, before the inclusion of the \emph{Introduction to process control} course in the curriculum, was near $55\,\%$. In the remaining control-related courses (\emph{Dynamic systems}, \emph{Instrumentation in control}, and \emph{Multi-variable systems}), the (first try) success rate was also over $70\,\%$.

In terms of the performance of the students in test, we stress that, in average, the overall grades in the \emph{Introduction to process control} course was positively affected when laboratory and practical experiments were conducted: the average of the final grades of the course exhibited an increase of approximately $20$\%. Also, the number of successful students after the inclusion of the laboratory also increased. 

\noindent\textbf{General comments}:\\ From a broad perspective, considering the students' testimonials and interviews, we can concretely infer that the \emph{Introduction to process control} classes has contributed in a positive way to the CAE \textcolor{red}{program} at UFSC. Many students report that having an early experience with control systems (already in their third semester) helped their learning process in this field of study. As an example, we highlight that all students who followed the first semester of  \emph{Introduction to process control} (in $2016$) were later admitted in their first try at \emph{Control systems} (in $2017$). We also emphasise that many students that do not take \emph{Introduction to process control} (coming from other universities or exchange programs) exhibited greater difficulties than their colleagues in other control courses.

\section{Conclusions and overall perspectives}
\label{secconc}

\noindent \textcolor{red}{In this paper, we debated a novel educational approach for control-theory in undergraduate engineering courses. In particular, we propose the inclusion of an introductory course on process control to the early semesters of undergraduate engineering programs. The focus of the proposed course is on the concepts of control theory themselves, and not on the mathematical tools that enable the more complex design analysis of control systems, as usually done. In this work, we detail the main guidelines of this introductory course, which should be taught pragmatically, based on experimental essays, and requiring, only, basic mathematical tools and knowledge of calculus and physics. }

\textcolor{red}{The main contribution of this work is not only the educational guideline, but rather on assessing how this approach has affected CAE undergraduate program at UFSC, taking into account the corresponding teaching-learning process from over seven years implementation. In particular, based on randomised interviews, we debate how the proposed course has helped the undergraduate students to learn control theory better, displaying better results in more advanced courses and serving to motivate them on the field. The vast majority of interviewed students explicitly mention highly positive views on the course and its motivational impacts along the graduation. }

\textcolor{red}{Our main finding relates to the fact that such \emph{Introduction to process control} course is, indeed, an interesting educational approach that can be successfully implemented in the first semesters of engineering programs. In particular, the implementation of the proposed introductory course at UFSC has had significant \textbf{practical results}: considering control and automation engineering or electrical engineering, for which it is intended to teach advanced control theory, our educational proposal has been validated as a motivational gateway in their respective undergraduate curricula.}

\textcolor{red}{In terms of perspectives of future results, we note that the proposed course can also serve to other engineering programs, for which the study of control system subjects are merely complementary or basic. Accordingly, we expect that the proposed course can be the single subject that addresses the theme of process control, given that its content covers the main subjects of practical implementation of controllers, and discusses on-off and PID control, which are the most used techniques in industrial practice. Furthermore, the proposal can be very interesting for technological degrees, which generally do not address mathematical tools in depth (such as Laplace transforms and $\mathcal{Z}$), but study control systems from a broader perspective.} 

\textcolor{red}{As a final message, we indicate that the complete syllabus of the proposed course is carefully detailed on our recent textbooks \citep{normey2021controle,normey2022controle}. Based on the detailed results from the teaching-learning process and the positive outcome of the course at UFSC, we \textbf{strongly advocate} that similar courses should be included to the curricula of other universities with undergraduate engineering program.}

\newpage
\backmatter

\bmhead{Supplementary information}

\section*{Declarations}

\subsection*{Acknowledgements}
\noindent An early version of paper was presented at \emph{XXIV Congresso Brasileiro de Autom\'atica} (CBA 2022).

\noindent The Authors thank the support from the Department of Automation and Systems at UFSC and the collaboration from Amir Naspolini and Gabriel Barbosa. 

\subsection*{Funding} 
\noindent The Authors acknowledge the financial support of CAPES, an entity of the Brazilian Government dedicated to the training of human resources, of the National Council for Scientific and Technological Development (CNPq, Brazil), under grants $304032/2019-0$ and $403949/2021-1$, 
and, also, the financial aid from the Human Resources Program of the National Agency of Petroleum, Natural Gas, and Bio-fuels (PRH-ANP), supported with resources from the investments of 
oil companies qualified in the Research, Development and Innovation Clause of ANP Resolution n$^{\text{o}}\, 50$/$2015$.

\subsection*{Conflicts of interest}
\noindent The Authors declare no conflict of interest. Both Authors contributed equally to this work.

\subsection*{Consent to participate} \noindent All undergraduate students consented to participate in the anonymous interview.

\subsection*{Code availability} 
\noindent \textcolor{red}{The open-source numerical simulator for simple control systems, as detailed in Section \ref{sec8b}, can be found in Google Drive\footnote{\url{https://drive.google.com/drive/folders/10TMMhUFv34BVOIA1vD_kQgAPrO1NwEGa?usp=sharing}} and GitHub\footnote{\url{https://github.com/GabrielBarbosaUFSC/FerramentasInterativas2023}}.}




\newpage
\bibliography{IntroControlebib}

\newpage
\appendix
\section{Control engineering \textcolor{red}{undergraduate programs} in Brazil}
\label{appendiceACursos}

\noindent Next, we briefly assess the control engineering undergraduate programs in Brazil, considering the seven surveyed universities presented in Table \ref{CAEnoBrasilTabela}. 

Accordingly, we present the curriculum of these seven CAE \textcolor{red}{undergraduate} programs, including a brief syllabus of each course and an overall assessment of the control engineering educational approach. 

In the following list, we use bold letters in order to mark the first subject that really discusses control system design (and not only the mathematical tools for analysis); again, we rank the universities according to the index discussed in \citep{morandin2020desempenho}:
\begin{itemize}
    \item \textbf{Universidade Estadual de Campinas (UNICAMP)}: 
    \begin{itemize}
        \item[($5^{\text{th}}$ sem.)] \emph{Applied mechanics}: Mathematical description of linear dynamical systems; transfer functions; Laplace transforms; system responses; stability criteria. vibration systems; free and forced vibrations; natural frequency and modes; experimental determination of moments of inertia, stiffness and damping.
         \item[($6^{\text{th}}$ sem.)] \emph{Mechanical systems}: Description, dimensioning, selection and use of mechanical elements; analysis of mechanical systems for energy conversion and transmission; mathematical modelling of dynamical systems.
          \item[($7^{\text{th}}$ sem.)] \textbf{\emph{Control of mechanical systems}}: Fundamental control concepts; basic control actions; frequency response; stability criteria and root locus; pole placement; notions of state and stability analysis.
           \item[($8^{\text{th}}$ sem.)] \emph{Advanced control systems}: Continuous and discrete processes and systems: modelling and process identification principles, dynamics, analysis and synthesis of feedback systems; industrial controllers and regulators; implementation of digital controllers; techniques and tools for analysis, simulation and design of industrial controllers.
            \item[($8^{\text{th}}$ sem.)] \emph{Instrumentation}: Basic instrumentation: multimeters, \textcolor{red}{analog} and digital oscilloscopes, digital logic analysers; static and dynamic characteristics of instruments and sensors; experimental data analysis; measurement and analysis of displacement, velocity, acceleration, force, torque, and mechanical power; problems in the amplification, transmission and storage of signals; sound measurements; pressure, flow and temperature measurements.
             \item[($9^{\text{th}}$ sem.)] \emph{Control systems laboratory}: Experiments related to the control of continuous and discrete systems, using industrial controllers and regulators.
              \item[($9^{\text{th}}$ sem.)] \emph{Data acquisition systems}: Data acquisition and conversion systems: A/D and D/A converters, frequency-voltage, current-voltage, sampling theorem, $\mathcal{Z}$-transform, digital filtering, discrete and fast discrete Fourier transform; design of digital filters; Architecture for digital signal processing: DSPs and ASICs
               \item[(Assessment)] In sum, there are seven control courses. Many aspects of modelling and instrumentation are considered within control system subjects. Modelling concerns are tied to stability aspects. Overall, the first notions of control design and implementation are only presented in the seventh semester (out of ten). Advanced and discrete-time topics are only discussed in the ninth semester.
    \end{itemize}
\end{itemize}

$$ \text{  }$$

    \begin{itemize}
    \item \textbf{Universidade Federal de Minas Gerais (UFMG)}:
 \begin{itemize}
        \item[($5^{\text{th}}$ sem.)] \emph{\textbf{Control engineering}}: Systems representations: differential equations, Laplace transforms, state-space representation, block diagrams and signal flow graph; first and second order systems; effects of a third pole and/or zero; damping coefficient estimation; Root locus; steady state error of feedback control systems; P, PI and PID controllers; Stability: Routh and Routh-Hurwitz criteria; Kharitonov's theorem; design of lead and lag controllers; minimum phase systems; gain margin and phase margin; Nyquist criterion and Nichols chart; design of frequency domain controllers. 
\item[($6^{\text{th}}$ sem.)] \emph{Digital Control}: Process control through computes; computer functions in process control; discrete-time systems and the $\mathcal{Z}$ transform: transform methods; theorems; solution of difference equations; signal flow graphs; representation via state variables; transfer functions; sampling and reconstruction of continuous-time signals: sampled control systems; the ideal sampler; conjugate transfer properties; sampling in open and closed-loop; relationship between $E(z)$ and $E^\star(s)$; modified $\mathcal{Z}$ transform; pure time delays; mapping of the $s$ plane on the $z$ plane; precision of control systems; stability: bilinear transformations, Routh-Hurwitz, Jury's and Nyquist criteria.
\item[($6^{\text{th}}$ sem.)] \emph{Instrumentation}: Measuring systems for open-loop applications; instrument operating modes; functional analysis
of instruments; error correction techniques; static and dynamic characterisation of instruments; signal analysis; electronic systems for
instrumentation; electrical coupling; Position, strain, pressure and flow sensors; temperature and level measurements.
\item[($7^{\text{th}}$ sem.)] \emph{Control laboratory I}: Integrate the technical-scientific knowledge acquired along the \textcolor{red}{program}, enabling students to implement real control applications. The processes are represented by real scaled, reduced plants, purchased on the market or developed in the laboratory. 
\item[($8^{\text{th}}$ sem.)] \emph{Control laboratory II}:
Implementation of digital controllers; programmable logic controllers; digital distributed control systems; real-time control techniques; numerical command systems.
\item[($8^{\text{th}}$ sem.)] \emph{Industrial process control techniques}: Rules and standards for instrumentation and control systems; advanced control strategies; controller tuning; notions of multi-variable control; predictive control and case studies.
\item[(Assessment)] A total of six control courses are available. The first course already contains aspects of designing basic controllers, altogether with the presentation of traditional frequency-domain techniques. The teaching is formally divided in continuous control and digital, discrete-time control systems.
\end{itemize}
$$ \text{  }$$
\end{itemize}
    \begin{itemize}
    \item \textbf{Universidade Federal de Santa Catarina (UFSC)}: 
    \begin{itemize}
        \item[($3^{\text{rd}}$ sem.)] \textbf{\emph{Introduction to process control}}\footnote{This course represents the novel educational approach detailed in this paper.}: Introduction to the process control problem: an intuitive overview of control systems, practical motivation and applications examples; basic concepts: static and dynamic models, manipulated and controlled variables, disturbances, operating point and regions, reference signals, feedback and feed-forward strategies; review and introduction of basic notions of physics and calculus applied to process control, notions of linearity and linearisation, minima and maxima of smooth functions; the basic control actions: proportional, integral and derivative actions; tuning controllers using empirical methods; practical examples and numerical simulations; other concepts and tools: implementation control through hierarchical layers, determining set-points and reference signals, tuning operating points based on economic or energy-related criteria. 
        \item[($4^{\text{th}}$ sem.)] \emph{Signals and linear systems}: Theory: introduction to continuous and discrete-time signals and linear systems; mathematical representation of linear systems; analysis of continuous-time and discrete linear time-invariant systems: response to initial conditions, impulse response, (continuous and discrete) convolution; interconnected systems, internal stability and BIBO stability, transient and permanent regimes; relations between the continuous and discrete-time domains; analysis of LTI systems using the Laplace and the $\mathcal{Z}$ transforms; inverse transforms; solution of differential and difference equations, transfer function, poles and zeros; Stability, influence of poles and zeros on the temporal response; Block algebra, application in feedback and control; polar and Bode diagrams, filtering; Fourier series and Transform; spectral decomposition; sampling and digital filtering; digital control of continuous-time systems, emulation of controllers and design principles in the $\mathcal{Z}$ domain. Laboratory: Experiments and simulations complementary content to theory, application of concepts in the analysis and simplified design of control systems and filters.
        \item[($5^{\text{th}}$ sem.)] \emph{Modelling and simulation of processes}: Industrial processes: phenomenological modelling, representation by block diagrams and instrumentation diagrams; mass and energy balance of chemical reaction processes; modelling with concentrated and distributed parameters; distillation columns; process simulation; use of academic and industrial simulators; practice: laboratory experiments with test-benches and industrial simulation software to study the behaviour of various types of processes (thermal, hydraulic, fermentation and distillation, etc.).
        \item[($6^{\text{th}}$ sem.)] \emph{Control systems}: Control structures; feedback control; stability and time response of feedback systems; The root locus (RL) design method as an analysis tool; frequency interpretation; Bode and polar diagrams and stability study; feedback control using the RL method (for continuous- and discrete-time systems); control with two degrees of freedom; pole-zero diagrams and their relation to closed-loop responses; Disturbance rejection in control systems; feedback and feed-forward control; cascade control; relational control; other process control structures and settings: buffer tank control, split range control, Smith predictor and filtered Smith predictor; industrial PID settings: configuration; structures; tuning methods; practical and operational aspects; implementation of digital controllers and coding aspects; introduction to robust control; Laboratory with simulation and experiments focused on the implementation of control strategies in micro-controllers, AD/DA and PC boards and with the use of industrial PIDs.
        \item[($6^{\text{th}}$ sem.)] \emph{Instrumentation in control}: Measurement: basic definitions involved in measurement systems; static and dynamic characteristics of measurement systems; specification and analysis of measurement systems for application in control systems; conditioning of measurement signals: two-, three- and four-wire electrical resistance measurement; Wheatstone bridge; resistive dividers and shunts; amplification; insulation; adjustment of input and output impedances; noise and interference; data acquisition: main types of data acquisition systems for instrumentation; sample-and-hold; A/D and D/A converters; operation: review of the main strategies used to drive loads (transistor as switch, PWM drive, H-bridge, proportional power amplifiers); digital controllers: implementation aspects and quantisation.
        \item[($7^{\text{th}}$ sem.)] \emph{Dynamic systems}: Linear and nonlinear dynamic systems; nonlinear problems in control engineering; state space representation and analysis (phase planes); qualitative analyses: equilibria, limit cycles and aperiodic behaviours; Hartman-Grobman linearisation theorem; structural stability; bifurcations; static nonlinearities in control systems (saturation, dead zone, hysteresis, backlash, friction, etc.); analysis methods in the frequency domain for detection of limit cycles; feedback systems with restrictions on control action: saturation and anti-windup methods; stability analysis of dynamical systems by the Lyapunov method; design techniques: (i) compensation of static nonlinearities ; (ii) linearisation by state and output feedback; (iii) design based on Lyapunov functions.
        \item[($8^{\text{th}}$ sem.)] \emph{Multi-variable systems}: Representation by state variables of continuous and sampled systems; analysis and design methods for multi-variable control systems; controllability and observability; canonical decomposition of linear systems; relationship between state representation and transfer function matrices; multi-variable poles and zeros; state feedback; observers; feedback control using estimated states; separation theorem; dynamic compensation.
        \item[(Assessment)] The \textcolor{red}{program} has eighth control-related subjects, considering that multi-variable control is an optional course for the undergraduates. The contact to control systems theory and aspects is early, as discussed along this paper. Along the courses, there is no formal separation of discrete- and continuous-time systems; all theoretical and implementation aspects are presented together. Notions of robust and advanced control are discussed in introductory courses.
    \end{itemize}
    \end{itemize}

    $$ \text{  }$$
    
    \begin{itemize}
    \item \textbf{Universidade Federal do Rio Grande do Sul (UFRGS)}:
    \begin{itemize}
        \item[($4^{\text{th}}$ sem.)] \emph{Modelling of mechanical systems}: The general objective  is to teach the undergraduate students how to develop a mathematical model that adequately represents the key aspects of a real physical system, typically in control applications; to this end, the activities follow the specific goal: (1) define the concept of model in the context of engineering and evaluate its importance; (2) identify the different types of systems modelling (iconic, mathematical, etc.), relating them to their most relevant applications; (3)develop mathematical models to represent the dynamic behaviour of physical systems of different natures (mechanical, thermal, electrical and hybrid systems); (4) study the most commonly used model representation techniques in the context of control engineering (Laplace transforms, block diagrams, state space representation); (5) use these tools to analytically predict the behaviour of the modelled systems, evaluating the most common patterns of their transient and permanent response; (6) simulate the behaviour of physical systems through appropriate computational packages. 
        \item[($5^{\text{th}}$ sem.)] \emph{Signals and systems}: Provide theoretical basis and analytical tools for the study of \textcolor{red}{analog} and digital systems and circuits; present these analytical tools in the context of engineering; introduce and develop the concept of frequency response of time-invariant linear systems; enable the undergraduate student to design simple filters; provide basic concepts of feedback systems, enabling the student to apply these concepts in analysis and design of electronic circuits, electro-mechanical devices, control systems and other applications.
    \item[($6^{\text{th}}$ sem.)] \emph{Fundamental instrumentation for control and automation}: Basic principles and methods of measuring physical quantities; metrology; main problems in measurement systems and know how to minimise them; transducers and their electro-electronic conditioners; development of experimental projects and practices in the field of instrumentation;
analysis of experimental data.
 \item[($6^{\text{th}}$ sem.)] \textbf{\emph{Control systems I}}: Basic notions of dynamic process modelling and identification of transfer function parameters from trials; feedback control strategies and the closed-loop effects; design, implementation and use of compensators for industrial mono-variable processes according to classical control theory, and the analysis of the effects of nonlinearities.
 \item[($7^{\text{th}}$ sem.)] \emph{Control systems II}: Analysis and design methods of control systems in the frequency domain; state-space representations; state feedback control laws based on state observers.
   \item[($7^{\text{th}}$ sem.)] \emph{Control laboratory}: reinforce and solidify, through experiments, the concepts and classical control techniques; apply modelling methods to physical process prototypes; apply controller design techniques to control physical process prototypes commonly found in the industry; observe and discuss the effects and phenomena caused by nonlinearities in control systems;identify the physical components of a control system; tune and implement single-loop and PLC control laws.
   \item[($8^{\text{th}}$ sem.)] \emph{Modelling and control of industrial processes}: Introduce the fundamental principles involved in the mathematical modelling of industrial processes, especially those
typically found in the chemical, petrochemical, refining, steel and bio-process industries; present the mathematical and computational methods suitable for simulating models of processes and operations of the
process industry, both in steady state and transient state; introduce the concepts and methods for the optimization of industrial processes, enabling the determination of the conditions
most suitable operational conditions for them; discuss the most suitable techniques for controlling typical industrial processes; introduce the principles of multi-variable control of industrial processes and, in particular, predictive control strategies,
enabling students to apply such techniques in industrial processes.
 \item[($8^{\text{th}}$ sem.)] \emph{Digital control systems}: Review the theoretical background on the analysis of sampled systems; provide the necessary theoretical foundations for the digital implementation of control systems; enable the student to design control algorithms; enable the student to identify mathematical models of processes necessary for the design of controllers; Discuss fundamental data-based control methodologies.
  \item[($9^{\text{th}}$ sem.)] \emph{Advanced process control}: Provide undergraduate students with theoretical and experimental studied on advanced industrial process control techniques, such as predictive control, state observers
applied to the development of virtual analysers, real-time optimization and methods for semi-autonomous system development; application to different types of industrial processes; production optimisation and cost reduction.
\item[(Assessment)] This \textcolor{red}{undergraduate} program contains the most number of control subjects (nine courses), focusing on the design and experimental application of diverse kinds of techniques. The degree covers several advanced techniques in depth. Nevertheless, the denser part of the control systems study happens after the first half of the \textcolor{red}{program}. The initial courses are focuses on modelling and mathematical tools.
    \end{itemize}
    \end{itemize}

    $$ \text{  }$$
    
    \begin{itemize}
    \item \textbf{Universidade Federal do Rio de Janeiro (UFRJ)}:
     \begin{itemize}
        \item[($4^{\text{th}}$ sem.)] \emph{Signals and systems}: Analysis techniques for continuous- and discrete-time systems; linear and time-invariant systems; transfer functions, series and Fourier transforms; Laplace and $\mathcal{Z}$ transforms; sampling theorem; solution of differential and difference equations using transforms; frequency response, Bode diagram and stability analysis; examples of applications with active filters: AM and FM modulation, feedback control implementation.
         \item[($5^{\text{th}}$ sem.)] \emph{Modelling dynamic systems}: Introduction to modelling; classification and types of models; mathematical modelling based on physical laws; mechanical, electrical, electro-mechanical, chemical systems, etc; experimental models; models based on neural networks; basic concepts of system identification; models of discrete event systems; introduction to computer simulation techniques.
          \item[($6^{\text{th}}$ sem.)] \textbf{\emph{Feedback control}}: Introduction to feedback control; principles of modelling dynamic systems;  linear and nonlinear models; linearisation; dynamic response: Laplace transform and transfer functions, block diagrams, experimental modelling; dynamic response compared to pole loci; basic feedback properties: disturbance rejection, sensitivity and dynamic reference tracking; PID control, stability: Routh criterion and BIBO stability; analysis and design by the root locus method; frequency response: Bode and Nyquist diagrams; design by the frequency method: Lead-Lag compensation.
           \item[($6^{\text{th}}$ sem.)] \emph{Process control and instrumentation}: Introduction to industrial instrumentation; feedback control systems and block diagram representations; industrial instrumentation in control loops; sensors and signal transmitters; controllers and stability of control loops; controller tuning methods; direct synthesis methods; feed-forward control systems; cascaded control.
            \item[($7^{\text{th}}$ sem.)] \emph{Advanced control}: Representation by state variables; analysis of state representations: canonical forms, stability, controllability, observability and dynamic responses; control by state feedback; pole placement and state estimators; compensator design using estimators; introduction to optimal control and LQR; Lyapunov stability; digital control: sampled systems, discrete-time system models and the $\mathcal{Z}$ transform; discrete-time design; state-space methods; sampling period choice.
             \item[($7^{\text{th}}$ sem.)] \emph{Signal processing}: Filtering; frequency response approximation; design of passive and active filters; frequency and impedance scaling and realisation; digital filters and their design: FIR and IIR.
              \item[($8^{\text{th}}$ sem.)] \emph{Nonlinear control systems}: Nonlinearities in control systems systems; phase plane analyses; stability points; limite cycles; stability and the Lyapunov method; absolute stability and the Popov and Circle criteria; harmonic linearisation and the descriptive function method; intentionally non-linear controllers.
               \item[(Assessment)] The CAE \textcolor{red}{program at} UFRJ is very similar to the one \textcolor{red}{at} UFSC. The course contains seven control-oriented courses; however, the first one specific to control systems is only taught in the sixth semester. Discrete- and continuous-time control are taught together, yet not introductory course to control theory is available.
    \end{itemize}
    \end{itemize}

    $$ \text{  }$$
    
    \begin{itemize}
    \item \textbf{Universidade Estadual Paulista (UNESP)}:
    \begin{itemize}
        \item[($5^{\text{th}}$ sem.)] \emph{Signals and systems}: Continuous and discrete time signals: power, energy, periodic, sinusoidal and exponential signals, impulse, step and ramp, signal construction; differential and difference equations, interconnection, memory, causality, stability, time invariance, linearity; linear and time invariant systems: convolution, step and impulse response, application examples; discrete-time Fourier transform: definition and examples, properties, convolution, inversion, applications in systems, transfer function; time- and frequency-domain characterisation of systems; sampling: mathematical representation, ideal reconstruction and zero order hold, aliasing, properties, pre-filter, notions of quantisation. $\mathcal{Z}$ transform: definition, region of convergence, properties, inverse transform, lead and lag operators, transfer functions; linear feedback systems: applications, analysis using root locus; stability criteria, phase and gain margins; discrete Fourier transform: definition, analysis and synthesis formulas, finite duration sequence, properties, discrete-time transform sampling, applications.
         \item[($6^{\text{th}}$ sem.)] \textbf{\emph{Control systems I}}: Fundamentals of control systems: transfer functions, open- and closed-loop transfer functions, block diagrams and operations; Analysis of control systems: transient and steady-state responses of first and second order systems; transient response performance specifications; steady-state response analysis; design via geometric root loci: effects of adding poles and zeros, compensation by cancellation, controller characteristics: proportional, proportional-integral, proportional-derivative, and PID controllers; characteristics of lead, lag and lead-lag compensators; frequency domain controller design: frequency response, Bode diagrams and project.
          \item[($6^{\text{th}}$ sem.)] \emph{Control laboratory I}: Simulation of dynamical systems; numerical integration; discrete-time approximation of continuous systems; Euler, trapezoidal and Runge-Kutta integrators; computer simulation softwares; block diagrams.
           \item[($7^{\text{th}}$ sem.)] \emph{Control systems II}: Linear systems in state-space: mathematical descriptions, transfer function matrices, linearisation and break-even points; linear Systems: canonical forms, Jordan form; stability: Lyapunov methods, input-output stability; controllability; observability; state feedback and state estimators.
           \item[($7^{\text{th}}$ sem.)] \emph{Control laboratory II}: Focus on test-bench experiments and digital control implementations; analyses of electrical and mechanical control systems through \textcolor{red}{analog} and/or digital controllers.
            \item[($8^{\text{th}}$ sem.)] \emph{Industrial process control}: Fundamentals and terminology of industrial process control systems; definition of terms used in the process control industry; basic process elements; dynamic characteristics of industrial processes; transfer and transport delay; analysis and design of industrial process control systems; performance criteria; industrial PID controllers; tuning methods; other techniques: feed-forward compensation, cascade control, dead-time compensators, and selective control; design and implementation of industrial process control systems through experimental essayes.
             \item[($8^{\text{th}}$ sem.)] \emph{Instrumentation and measurement systems}: Measurement concepts, instruments and techniques, measurement systems and units in used in the International System; measurement standards and concepts of calibration and traceability; signal conditioning, amplification and filtering; AD and DA conversion techniques; displacement and velocity sensors (potentiometer, LVDT, RVDT, encoder, tachogenerator, extensometer); acceleration sensors (piezoelectric, accelerometer); force, torque and pressure sensors (strain gauge, Piezoelectric, Pitot tube); temperature sensors (thermo resistor, thermistor, thermocouple, pyrometer); flow sensors (Pitot tube, anemometer, drag, rotameter, orifice plate, mouthpiece, Venturi); magnetic field sensors (Hall effect); pressure, force, displacement, temperature, flow and level measurement.  valves and final control elements: solenoid, electro-pneumatic and electro-hydraulic valves; process and instrumentation diagrams.
    \item[(Assessment)] Seven control-related courses, with similar emphasis to continuous and discrete-time systems. Mathematical tools are presented first, considering digital and \textcolor{red}{analog} control. Then, only in the sixth semester, the undergraduate students are present to design methods and controllers, with experimental essays done together.
    \end{itemize}
    \end{itemize}

    $$ \text{  }$$
    
    \begin{itemize}
    \item \textbf{Universidade Federal de Pernambuco (UFPE)}:
    \begin{itemize}
         \item[($6^{\text{th}}$ sem.)] \textbf{\emph{Control engineering I}}: Introduction to control systems and presentation of the concept of open-loop and closed-loop control; mathematical modelling and analysis of dynamic systems in time- and frequency-domain as an initial step to the design of control systems; use of computational tools for analysis of dynamical systems.
          \item[($6^{\text{th}}$ sem.)] \emph{Industrial instrumentation}: Characterisation of displacement, force, torque, pressure, flow, level, temperature and thermal flow transducers; calibration of transducers; instrumentation, measurement, recording and handling of experimental data; automation and measurement.
           \item[($7^{\text{th}}$ sem.)] \emph{Control engineering II}: Design of continuous-time \textcolor{red}{analog} control systems by pole placement, heuristic methods, root locus and frequency responses, considering transient and stationary specification; system stability analysis using the Routh-Hurwitz and Nyquist criteria.
            \item[($7^{\text{th}}$ sem.)] \emph{Control engineering laboratory}: Obtaining a dynamic model from experimental tests; tuning of controllers using heuristic methods; characterisation of open-loop and closed-loop systems; design and implementation of signal conditioning circuits; implementation of classic control strategies.
             \item[($8^{\text{th}}$ sem.)] \emph{Digital control}: Introduction to digital control systems; $\mathcal{Z}$ transform; sampling, data retention and reconstruction of continuous signals from samples; open-loop and closed-loop discrete-time systems; correspondence between the $s$ and the $z$ plane; stability criteria; transient and steady-state response analysis; design of digital controllers based on discretisation of \textcolor{red}{analog} controllers; direct methods of designing digital controllers.
               \item[(Assessment)] In this CAE \textcolor{red}{undergraduate program}, there are fewer control-related courses (only six), with focus on classical topics of feedback control systems and their implementation. The first course, in the sixth semester, presents the mathematical tools (frequency analysis and root loci) and the design of controllers together.
    \end{itemize}
\end{itemize}

\newpage
\section{Course syllabus}
\label{appendixSyllabus}

\noindent Next, we present the detailed syllabus of the proposed \emph{Introduction to process control} course, to be taught just after basic calculus and physics courses:
\begin{itemize}
    \item Part 1: Processes and systems
    \begin{itemize}
        \item Topic 1: Processes
        \begin{itemize}
            \item What are processes?
            \item Block diagrams and basic operations
            \item Process control: a general insight
            \item Why to use control systems and feedback? 
            \item Control design specifications
            \item Understanding control systems: actuators and sensors; controlled variables, process variables and disturbances
            \item Motivational examples from our daily routine
            \item Mono-variable and multi-variables schemes
            \item Closed-loop control and automatic operation
        \end{itemize}
        \item Topic 2: Signals and systems
        \begin{itemize}
            \item System representations and models
            \item Different kinds of models and diagrams
            \item What are signals?
            \item Signals: representations and use
            \item Useful signals: step, ramp, unitary pulse, sines and cosines, exponential terms
            \item Signals: energy and power
            \item Operations and transformation of signals
            \item Sampling, interpolation, and quantisation
            \item Digital signal processing
        \end{itemize}
        \item Topic 3: System properties
        \begin{itemize}
            \item Classifications:  continuous and discrete, causal and non-causal, time-invariant and time-varying, linear and nonlinear, stable and unstable
            \item Examples and models of each type of system
            \item Linearity and properties of linear systems
            \item Stability of equilibria: an intuitive definition and formal characterisation
            \item Operational point and working regions
            \item Stability in the neighbourhood of a steady-state point
            \item The concept of BIBO stability
        \end{itemize}
    \end{itemize}
    \item Part 2: Models and time responses
    \begin{itemize}
        \item Topic 4: Modelling
        \begin{itemize}
            \item Basic principles of modelling dynamic systems
            \item Examples of continuous-time and discrete-time phenomenological models
            \item Process analyses: static and dynamic models
            \item Static characteristics
            \item Transient and steady-state regimes
            \item Process simulation using numerical software
            \item Tuning first-order models from experimental data 
            \item Approximate linear models of nonlinear dynamic systems
            \item First-order Taylor linearisation
        \end{itemize}
         \item Topic 5: Time response of first-order linear systems
        \begin{itemize}
            \item An intuitive approach to time responses
            \item Continuous-time systems and ordinary differential equations (ODEs)
            \item Solution to an ODE: stable and unstable cases
            \item Natural and forced responses
            \item Solution to an ODE: integrator case
            \item Discrete-time systems and difference equations (DEs)
            \item General solution to DEs from geometric progressions
            \item Time response of sampled systems
        \end{itemize}
         \item Topic 6: Closing the loop: dynamic output feedback control
        \begin{itemize}
            \item The different control objectives
            \item Hierarchical control and the automation pyramid
            \item Different control layers and their respective tasks
            \item Closed-loop control and why to use feedback
            \item Control design specifications: set-point tracking, disturbance rejection, settling time, operational constraints
            \item Control within operational ranges 
            \item Approximate models including transport delay
            \item Basic notions of system identification
            \item Practical aspects: noise and variable normalisation
        \end{itemize}
    \end{itemize}
    \item Part 3: Process control
    \begin{itemize}
        \item Topic 7: on-off control
        \begin{itemize}
            \item The basic idea of on-off control
            \item Implementation aspects in continuous- and discrete-time
            \item on-off control using \textcolor{red}{analog} relays
            \item Hysteresis curve and closed-loop diagrams
            \item System responses under on-off control in continuous- and discrete-time
            \item Computing periods of time in on and off conditions
            \item Analyses in the presence of disturbances
        \end{itemize}
         \item Topic 8: Proportional control
        \begin{itemize}
           \item An intuitive notion of a proportional control action
            \item Implementing proportional control: positive and negative gains
            \item Proportional range and steady-state analyses
            \item The use of a bias signal for steady-state set-point tracking
            \item Implementation aspects: \textcolor{red}{analog} electronic circuit and digital embedding
            \item Practical examples
            \item The closed-loop solution: stable systems
            \item The closed-loop solution: unstable and integrator systems
            \item The additional bias signal for steady-state set-point tracking
            \item An automatic bias signal 
            \item Disturbance rejection effects and insights on integral action
        \end{itemize}
         \item Topic 9: Proportional-integral control
        \begin{itemize}
             \item An intuitive notion of why to use integral action in feedback 
            \item Motivational examples of integral action
            \item Integral control 
             \item Null steady-state error condition
            \item Integral control compared to an automatic bias 
            \item Tuning integral controllers: analysis of the transient regime
            \item PI control: weighted sum of both control actions
            \item Implementing PI control: \textcolor{red}{analog} and digital aspects
            \item Tuning PI controllers: basic rules for delay-free systems
            \item Tuning PI controllers: Ziegler-Nichols rules
            \item Tuning PI controllers: Skogestad rules
        \end{itemize}
         \item Topic 10: PI control and practical aspects
        \begin{itemize}
            \item Debating practical problems of PI control using realistic examples
            \item Set-point weighting strategies
            \item Switching from automatic to manual control modes
            \item Input saturation and anti-windup strategies
        \end{itemize}
         \item Topic 11: Proportional-integral-derivative control
        \begin{itemize}
            \item Derivative control as a prediction of future dynamics
            \item PID controllers and implementation structures
            \item Tuning PID controllers: open-loop Ziegler-Nichols rules
            \item Tuning PID controllers: closed-loop Ziegler-Nichols rules
            \item Tuning PID controllers: IMC rules
            \item Implementing PID controllers: \textcolor{red}{analog} and digital aspects
            \item Measurement noise and the resulting effects
            \item Noise filtering in control systems
            \item Implementing low-pass measurement filters
            \item Noise filtering in PID controllers
            \item Feed-forward control
        \end{itemize}
    \end{itemize}
\end{itemize}





\end{document}